\documentclass[fleqn,oldversion]{aa}
\usepackage{graphicx}
\usepackage{amsmath,amsbsy}
\usepackage[varg]{txfonts}
\usepackage{setspace}
\usepackage{natbib}
\newcommand{\se}[1]{\mbox{Sect.\ \ref{sec:#1}}}

\newcommand{\Se}[1]{\mbox{Section\ \ref{sec:#1}}}
\newcommand{\eq}[1]{\mbox{Eq.\ (\ref{eq:#1})}}
\newcommand{\eqs}[2]{Eqs.\ (\ref{eq:#1}) and (\ref{eq:#2})}

\newcommand{\Eqs}[2]{Equations\ (\ref{eq:#1}) and (\ref{eq:#2})}
\newcommand{\fg}[1]{\mbox{Fig.\ \ref{fig:#1}}}
\newcommand{\fgs}[2]{Figs.\ \ref{fig:#1} and \ref{fig:#2}}
\newcommand{\Fg}[1]{\mbox{Figure\ \ref{fig:#1}}}
\newcommand{\Tb}[1]{\mbox{Table\ \ref{tab:#1}}}
\newcommand{\app}[1]{\mbox{Appendix\ \ref{app:#1}}}
\newcommand{\ie}{i.e.,}
\newcommand{\vs}{vs.}
\newcommand{\eg}{e.g.,}

\newcommand{\cf}{cf.}
\newcommand{\micr}{\ensuremath{\mu\mathrm{m}}}
\newcommand{\sumi}{(i)}
\newcommand{\sumii}{(ii)}
\newcommand{\sumiii}{(iii)}

\def\figwidth{88mm} 
\begin{document}
\title{Dust coagulation and fragmentation in molecular clouds}
\subtitle{I. How collisions between dust aggregates alter the dust size distribution}
\author{C.W. Ormel \inst{1,2} \and
        D. Paszun \inst{3} \and
        C. Dominik \inst{3,4} \and
        A.G.G.M. Tielens \inst{5,6}
        }
\institute{Kapteyn Astronomical Institute, University of Groningen, PO box 800, 9700 AV  Groningen, The Netherlands
          \and
          Max-Planck-Institut f\"ur Astronomie, K\"onigstuhl 17, 69117, Heidelberg, Germany;
          \email{ormel@mpia-hd.mpg.de}
          \and
          Sterrenkundig Instituut ‘Anton Pannekoek’, Kruislaan 403, 1098 SJ Amsterdam, The Netherlands; 
          \email{C.Dominik@uva.nl}
          \and
          Afdeling Sterrenkunde, Radboud Universiteit Nijmegen, Postbus 9010, 6500 GL Nijmegen, The Netherlands
          \and
          Ames Research Center, NASA, Mail Stop 245-3, Moffett Field, CA 94035, USA
          \and
          Leiden Observatory, Leiden University, P.O. Box 9513, 2300 RA Leiden, The Netherlands\\
          \email{tielens@strw.leidenuniv.nl}
          }
\abstract{The cores in molecular clouds are the densest and coldest regions of the interstellar medium (ISM).  In these regions ISM-dust grains have the potential to coagulate.  This study investigates the collisional evolution of the dust population by combining two models: a binary model that simulates the collision between two aggregates and a coagulation model that computes the dust size distribution with time.  In the first, results from a parameter study quantify the outcome of the collision -- sticking, fragmentation (shattering, breakage, and erosion) -- and the effects on the internal structure of the particles in tabular format.  These tables are then used as input for the dust evolution model, which is applied to an homogeneous and static cloud of temperature 10 K and gas densities between $10^3$ and $10^7\ \mathrm{cm^{-3}}$.  The coagulation is followed locally on timescales of $\sim$$10^7\ \mathrm{yr}$.  We find that the growth can be divided into two stages: a growth dominated phase  and a fragmentation dominated phase.  Initially, the mass distribution is relatively narrow and shifts to larger sizes with time.  At a certain point, dependent on the material properties of the grains as well as on the gas density, collision velocities will become sufficiently energetic to fragment particles, halting the growth and replenishing particles of lower mass.  Eventually, a steady state is reached, where the mass distribution is characterized by a mass spectrum of approximately equal amount of mass per logarithmic size bin. The amount of growth that is achieved depends on the cloud's lifetime.  If clouds exist on free-fall timescales the effects of coagulation on the dust size distribution are very minor.  On the other hand, if clouds have long-term support mechanisms, the impact of coagulation is important, resulting in a significant decrease of the opacity on timescales longer than the initial collision timescale between big grains.}
\titlerunning{Dust Coagulation and Fragmentation in Molecular Clouds I.}
\keywords{ISM: dust, extinction -- ISM: clouds -- Turbulence -- Methods: numerical}
\maketitle
\section{Introduction}
\label{sec:intro}
Dust plays a key role in molecular clouds. Extinction of penetrating FUV photons by small dust grains allows molecules to survive. At the same time, gas will accrete on dust grains forming ice mantles consisting of simple molecules \citep{1982A&A...114..245T,1992ApJS...82..167H}. Moreover, surface chemistry provides a driving force towards molecular complexity  \citep{1992ApJ...399L..71C,2008ApJ...674..984A}.  Finally, dust is often used as a proxy for the total gas (H$_2$) column density, either through near-IR extinction measurements or through sub-millimeter emission studies \citep{2006ApJ...653..383J,2007A&A...462L..17A,JorgensenEtal:2008}. Dust is often preferred as a tracer in these types of studies because it is now well established that -- except for pure hydrides -- all species condense out in the form of ice mantles at the high densities of prestellar cores \citep{2006A&A...456..215F,2007ARA&A..45..339B,2007A&A...462..221A}.  Thus, it is clear that our assessment of the molecular contents of clouds, as well as the overall state of the star and planet formation process, are tied to the properties of the dust grains -- in particular, its size distribution.

The properties of interstellar dust are, however, expected to change during its sojourn in the molecular cloud phase.  First, condensation from the gas phase causes grain sizes to increase, forming ice mantles. This growth is limited, however, because there are many small grains -- which dominate the total grain surface area -- over which the ice should be distributed; if all the condensible gas freezes out, the thickness of the ice mantles is still only 175\ \AA\ \citep{1985prpl.conf..621D}. Therefore, in dense clouds, coagulation is potentially a much more important promoter of dust growth. On a long timescale ($>$$10^8\ \mathrm{yr}$), the interstellar grain size distribution is thought to reflect a balance between coagulation in dense clouds and shattering in interstellar shocks as material constantly cycles between dense and diffuse ISM phases \citep{JonesEtal:1996,DominikTielens:1997}. 

Infrared and visual studies of the wavelength dependence of linear polarization and the ratio of total-to-selective extinction were among the first observational indications of the importance of grain growth in molecular clouds \citep{CarrascoEtal:1973,WilkingEtal:1980,Whittet:2005}. Early support for grain growth by coagulation in molecular clouds was also provided by a Copernicus study that revealed a decreased amount of visual extinction per H-nucleus in the $\rho$-Oph cloud relative to the value in the diffuse interstellar medium \citep{Jura:1980}. These type of visual and UV studies are by necessity limited to the outskirts of molecular clouds. Subsequent IR missions provided unique handles on the properties of dust deep inside dense clouds. In particular, comparison of far-IR emission maps taken by IRAS and Spitzer and near-IR extinction maps derived from 2MASS indicate grain growth in the higher density regions \citep{SchneeEtal:2008}. Likewise, evidence for grain coagulation is also provided by a comparison of visual absorption studies (e.g., star counts) and sub-millimeter emission studies which imply that the smallest grains have been removed efficiently from the interstellar grain size distribution \citep{StepnikEtal:2003}. Similarly, a recent comparison of Spitzer-based, mid-IR extinction and submillimeter emission studies of the dust characteristics in cloud cores reveals systematic variations in the characteristics as a function of density consistent with models of grain growth by coagulation \citep{ButlerTan:2009}. Dust-to-gas ratios derived from a comparison of line and continuum sub-millimeter data is also consistent with grain growth in dense cloud cores \citep{GoldsmithEtal:1997}. In recent years, X-ray absorption studies with Chandra have provided a new handle on the total H column along a line of sight -- that can potentially probe much deeper inside molecular clouds than UV studies -- and in combination with Spitzer data, the decreased dust extinction per H-nucleus reveals grain growth in molecular clouds \citep{WinstonEtal:2007}. Finally, Spitzer/IRS allows studies of the silicate extinction inside dense clouds and a comparison of near-IR color excess with $10\ \mu\mathrm{m}$ optical depth reveals systematic variations which is likely caused by coagulation \citep{ChiarEtal:2007}. This is supported by an analysis of the detailed absorption profile of the 10 $\mu$m silicate absorption band in these environments \citep{BoweyEtal:1998,vanBreemenEtal:2009}. 

Because it is the site of planet formation, theoretical coagulation studies have largely focused on grain growth in protoplanetary disks \citep{1993prpl.conf.1031W}.  In molecular clouds, dust coagulation has been theoretically modeled by \citet{Ossenkopf:1993} and \citet{WeidenschillingRuzmaikina:1994}.  In these studies, coagulation is driven by processes that provide grains with a relative motion. For larger grains ($\gtrsim$100 \AA) turbulence in particularly is important in providing relative velocities.  These motions -- and the outcomes of the collisions -- are very sensitive to the coupling of the particles to the turbulent eddies, which is determined by the surface area-to-mass ratio of the dust particles.  At low velocities, grain collisions will lead to the growth of very open and fluffy structures, while at intermediate velocities compaction of aggregates occurs.  At very high velocities, cratering and catastrophic destruction will halt the growth \citep{DominikTielens:1997,PaszunDominik:2009,BlumWurm:2008}.  Thus, to study grain growth requires us to understand the relationships between the macroscopic velocity field of the molecular cloud, the internal structure of aggregates (which follows from its collision history), and the microphysics of dust aggregates collisions.  In view of the complexity of the coagulation process and the then existing, limited understanding of the coagulation process, previous studies of coagulation in molecular cloud settings have been forced to make a number of simplifying assumptions concerning the characteristics of growing aggregates.

Theoretically, our understanding of the coagulation process has been much improved by the development of the atomic force microscope, which has provided much insight in the binding of individual monomers. This has been translated into simple relationships between velocities and material parameters, which prescribe under which conditions sticking, compaction, and fragmentation occur \citep{1993ApJ...407..806C,DominikTielens:1997}.  Over the last decade, a number of elegant experimental studies by Blum and coworkers \citep[see][]{BlumWurm:2008} have provided direct support for these concepts and in many ways expanded our understanding of the coagulation process. Numerical simulations have translated these concepts into simple recipes, which link the collisional parameters and the aggregate properties to the structures of the evolving aggregates \citep{PaszunDominik:2009}. Together with the development of Monte Carlo methods, in which particles are individually followed \citep{OrmelEtal:2007,ZsomDullemond:2008}, these studies provide a much better framework for modeling the coagulation process than hitherto possible.  

In this paper, we reexamine the coagulation of dust grains in molecular cloud cores in the light of this improved understanding of the basic physics of coagulation with a two-fold goal. First, we will investigate the interrelationship between the detailed prescriptions of the coagulation recipe and the structure, size, and mass of aggregates that result from the collisional evolution.  Therefore, a main goal of this work is to explore the full potential of the collision recipes, \eg\ by running simulations that last long enough for fragmentation phenomena to become important.  Second, we aim to give a simple prescriptions for the temporal evolution of the total grain surface area in molecular clouds, thereby capturing its observational characteristics, in terms of the physical conditions in the core. 

This paper is organized as follows. \Se{struc} presents a simplified and static model of molecular clouds we adopted in our calculations.  \Se{colmod} describes the model to treat collisions between dust grains and, more generally, aggregates of dust grains.  In \se{results} the results are presented: we discuss the imprints of the collision recipe on the coagulation and also present a parameter study, varying the cloud gas densities and the dust material properties. In \se{discus} we review the implications of our result to molecular clouds and \se{concl} summarizes the main conclusions.

\section{Density and velocity structure of molecular clouds}
\label{sec:struc}
The physical structure of molecular clouds -- the gas density and temperature profiles -- is determined by its support mechanisms: thermal, rotation, magnetic fields, or turbulence.  If there is only thermal support to balance the cloud's self-gravity and the temperature is constant, the density, assuming spherical symmetry, is given by $\rho_\mathrm{g} \propto r^{-2}$, where $\rho_\mathrm{g}$ is the gas density and $r$ the radius.\footnote{A list of symbols is provided in \app{appD}.}  However, the isothermal sphere is unstable as it heralds the collapse phase \citep{1977ApJ...214..488S}.  The cloud then collapses on a free-fall timescale
\begin{equation}
  t_\mathrm{ff} = \sqrt{\frac{3\pi}{32G\rho_\mathrm{g}}} = 1.1\times10^5\ \mathrm{yr}\ \left( \frac{n}{10^5\ \mathrm{cm^{-3}}} \right)^{-1/2},
  \label{eq:tff}
\end{equation}
where $G$ is Newton's gravitational constant, $n=\rho_\mathrm{g}/m_\mathrm{H}\mu$ the number density of the molecular gas,\footnote{\label{foot:gasdensity}Note that our definition of density -- $n$, the number of gas molecules per cm$^3$ -- differs from the density of hydrogen nuclei ($n_\mathrm{H}$), which is more commonly referred to as density in the dust/ISM communities. For cosmic abundances $n_\mathrm{H}$ is related to $n$ as $n\simeq1.7n_\mathrm{H}$. } $m_\mathrm{H}$ the hydrogen mass, and $\mu=2.34$ the mean molecular mass.  Thermally supported cores are stable if the thermal pressure wins over gravity, a situation described by the Bonnor-Ebert sphere, where an external pressure confines the cloud (still assuming a constant temperature). 

Magnetic fields in particular are important to support the cloud against the opposing influence of gravity. Ion-molecule collisions provide friction between the ions and neutrals and in that way couple the magnetic field to the neutral cloud. Over time the magnetic field will slowly leak out on an ambipolar diffusion timescale

\begin{equation}
  t_\mathrm{ad} \sim \frac{3 K_{in}}{4\pi\mu m_\mathrm{H} G} \left( \frac{n_i}{n} \right) \simeq 3.7\times10^6\ \mathrm{yr} \left( \frac{n}{10^5\ \mathrm{cm^{-3}}} \right)^{-1/2},
  \label{eq:tad}
\end{equation}
where $K_{in}$ is the ion-molecular collision rate and $n_i$ the density in ions. In \eq{tad} we have used an ion-neutral collision rate of $K_{in}=2\times10^{-9}\ \mathrm{cm^3\ s^{-1}}$ and a degree of ionization due to cosmic rays of $n_i/n=2\times10^{-5}/\sqrt{n}$ \citep{Tielens2005}.

Turbulence is another possible support mechanism of molecular cores, but its nature is dynamic -- rather than (quasi)static.  At large scales it provides global support to molecular clouds, whereas at small scales it locally compresses the gas. If these overdensities exist on timescales of \eq{tff}, collapse will follow.  This is the gravo-turbulent fragmentation picture of turbulence-dominated molecular clouds, where the (supersonic) turbulence is driven at large scales, but also reaches the scales of quiescent (subsonic) cores \citep{2004RvMP...76..125M,2005ApJ...620..786K}. In this dynamical, turbulent-driven picture both molecular clouds and cores are transient objects.

Thus, cloud cores will dynamically evolve due to either ambipolar diffusion and loss of supporting magnetic fields or due to turbulent dissipation, or simply because the core is only a transient phase in a turbulent velocity field.  In this work, for reasons of simplicity, we consider only a static cloud model -- the working model -- in which turbulence is unimportant for the support of the core, but where (subsonic) turbulence is included in the formalism for the calculation of relative motions between the dust particles.

\subsection{Working model}
\label{sec:working-model}
In this exploratory study we will for simplicity adopt an homogeneous core of mass given by the critical Jeans mass.  Moreover, we assume the cloud is turbulent, but neglect the influence of the turbulence on the support of the cloud.  Thus, our approximation is probably applicable for high density, low mass cores as velocity dispersions increase towards high mass cores \citep{1998ApJS..117..387K}. The homogeneous structure ensures that collision timescales are the same throughout the cloud, \ie\ the coagulation and fragmentation can be treated locally.  In our calculations, the sensitivity of the coagulation on the gas density $n$ will be investigated and the relevant coagulation and fragmentation timescales will be compared to the timescales in Eqs.\ (\ref{eq:tff}) and (\ref{eq:tad}).

Starting from the isodense sphere, the cloud outer radius is given by the Jeans length \citep{1987gady.book.....B}
\begin{equation}
  L_\mathrm{J} = \frac{1}{2}\sqrt{\frac{\pi c_\mathrm{g}^2}{G \rho_\mathrm{g}}} = 0.033\ \mathrm{pc}\ \left( \frac{n}{10^5\ \mathrm{cm^{-3}}} \right)^{-1/2} \left( \frac{T}{10\ \mathrm{K}} \right)^{1/2},
\end{equation}
where $c_\mathrm{g} = kT/\mu$ is the isothermal sound speed and $T$ the temperature of the cloud. A temperature of $10\ \mathrm{K}$ is adopted.  The sound crossing time $L_\mathrm{J}/c_\mathrm{g}$ is comparable to the free-fall time of the cloud.

Regarding the driving scales of the turbulence we assume \sumi\ that the largest eddies decay on the sound crossing time, \ie\ $t_\mathrm{L}= L_\mathrm{J}/c_\mathrm{g}$, and \sumii\ that the fluctuating velocity at the largest scale is given by the sound speed, $v_\mathrm{L}=c_\mathrm{g}$. Thus, the turbulent viscosity is $\nu_\mathrm{t} = L v_\mathrm{L} = v_\mathrm{L}^2 t_L = c_\mathrm{g} L_\mathrm{J}$ with $L=L_\mathrm{J}$ the size of the largest eddies. Although our parametrization of the large eddy quantities seems rather ad-hoc, we can build some trust in this relation by considering the energy dissipation rate $v_\mathrm{L}^3/L$ per unit mass, which translates into a heating rate of
\begin{equation}
  n\Gamma = \frac{v_L^3}{L} \rho_\mathrm{g} = 2.5\times10^{-23}\ \mathrm{erg\ cm^{-3}\ s^{-1}} \left( \frac{T}{10\ \mathrm{K}} \right) \left( \frac{n}{10^5\ \mathrm{cm^{-3}}} \right)^{3/2}.
  \label{eq:heat}
\end{equation}
Based upon observational studies of turbulence in cores, \citet{Tielens2005} gives a heating rate of $n\Gamma = 3\times10^{-28}n\ \mathrm{erg\ s^{-1}}$, with which \eq{heat} reasonably agrees for the range of densities we consider.  Additionally, the adoption of the crossing time and sound speed for the large eddy properties are natural upper limits. 

The turbulent properties further follow from the Reynolds number, 
\begin{equation}
  \label{eq:Re}
  \mathrm{Re} = \frac{\nu_\mathrm{t}}{\nu_\mathrm{m}} =  \frac{v_\mathrm{L}L}{c_\mathrm{g} \ell_\mathrm{mfp}/3} = 6.2\times10^7\ \left( \frac{n}{10^5\ \mathrm{cm^{-3}}} \right)^{1/2} \left( \frac{T}{10\ \mathrm{K}} \right)^{1/2},
\end{equation}
where $\nu_\mathrm{m}$ is the molecular viscosity and $\ell_\mathrm{mfp}$ the mean free path of a gas particle.  Assuming a Kolmogorov cascade, the turn-over time and velocity at the inner scale follow from the Reynolds number
\begin{subequations}
\begin{equation}
 \label{eq:ts}
 t_\mathrm{s} = \mathrm{Re}^{-1/2} t_\mathrm{L} = 2.2\times10^1\ \mathrm{yr}\ \left( \frac{n}{10^5\ \mathrm{cm\ s^{-1}}} \right)^{-3/4} \left( \frac{T}{10\ \mathrm{K}} \right)^{-1/4};
\end{equation}
\begin{equation}
  \label{eq:vs}
  v_\mathrm{s} = \mathrm{Re}^{-1/4} v_\mathrm{L} = 2.1\times10^2\ \mathrm{cm\ s^{-1}}\ \left( \frac{n}{10^5\ \mathrm{cm\ s^{-1}}} \right)^{-1/8} \left( \frac{T}{10\ \mathrm{K}} \right)^{3/8}.
\end{equation}
\end{subequations}

A collisional evolution model requires a prescription for the \textit{relative} velocities $\Delta v$ between two solid particles.  Apart from turbulence, other mechanisms, reflecting differences in the thermal, electrostatic, and aerodynamic properties of particles, will also provide particles with a relative motion. However, under most molecular cloud conditions turbulence will dominate the velocity field \citep{Ossenkopf:1993} and in this work we only consider turbulence.  Then, the surface area-to-mass ratio of the dust particles is of critical importance since this quantity determines the amount of coupling between the dust particles and the gas. We use the analytic approximations of \citet{OrmelCuzzi:2007} for the relative velocity between two particles. These expressions only include contributions that arise as a result of the particle's inertia in a turbulent velocity field and do not contain contributions that arise from gyroresonance acceleration \citep[see, \eg][]{YanEtal:2004}. See \app{app1} for more details.

\section{Collision model}
\begin{figure}
  \centering
  \includegraphics[width=0.4\textwidth,clip]{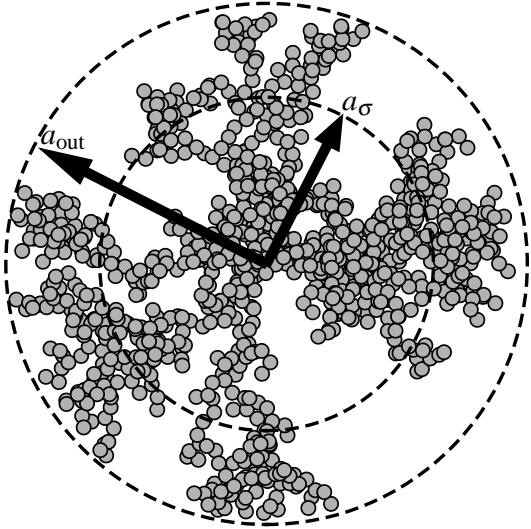}
  \caption{\label{fig:fractal}2D projection of a fluffy aggregate with indication of the geometrical radii, $a_\sigma$, and the outer radii, $a_\mathrm{out}$.}
\end{figure}
\label{sec:colmod}
Dust grains that collide can stick together, forming aggregates (see \fg{fractal}). In this work we consider the collisional evolution of the distribution of aggregates, $f(N,t)$: the number of aggregates of mass $N$ with time.  Many works have studied aggregate growth  under the conditions of perfect sticking upon contact, neglecting the effects of the impact energy on the structure of aggregates \citep{Meakin1988,MeakinDonn:1988,Ossenkopf:1993}. Of special interest are the particle-cluster aggregation (PCA) and cluster-cluster aggregation (CCA) modes. In PCA, the aggregates collide only with single grains, while in CCA the collision partners are of similar size and structure. In CCA, the emerging structures become true fractals, with a fractal dimension $\sim$2. For PCA, however, an homogeneous structure is eventually reached at a filling factor of $\sim$15\%.

However, the assumption that the internal structure is fixed (as in fractals) becomes invalid if the collisions take place between particles of different size. Furthermore, at higher energies the assumption of `hit-and-stick' breaks down: aggregate bouncing, compaction (in which the constituent grains rearrange themselves), and fragmentation lead to a rearrangement of the internal structure.  These collisional processes, except bouncing, are included in our collision model.

We quantify the internal structure of aggregates in terms of the geometrical filling factor, $\phi_\sigma$, defined as
\begin{equation}
  \phi_\sigma = N \biggl( \frac{a_0}{a_\sigma} \biggr)^3,
  \label{eq:phi-geom-def}
\end{equation}
where we have assumed that the aggregate contains $N$ equal size grains of radius $a_0$ with $a_\sigma=\sqrt{\sigma/\pi}$ the projected surface equivalent radius of the aggregate.  For very fluffy aggregates $a_\sigma$ can be much less than the outer radius of the aggregate, $a_\mathrm{out}$, see \fg{fractal}. The definition of the filling factor in terms of the projected area determines its aerodynamic behavior, and thereby the relative velocities ($\Delta v$) between the dust aggregates.\footnote{The compactness parameter $\phi_\sigma$ is the inverse of the enlargement factor $\psi$, previously used in \citet{OrmelEtal:2007}. \citet{Ossenkopf:1993} uses $x=\phi_\sigma^{-2}$ as its structural parameter.} 
\begin{figure*}
  \centering
  \includegraphics[width=\textwidth,clip]{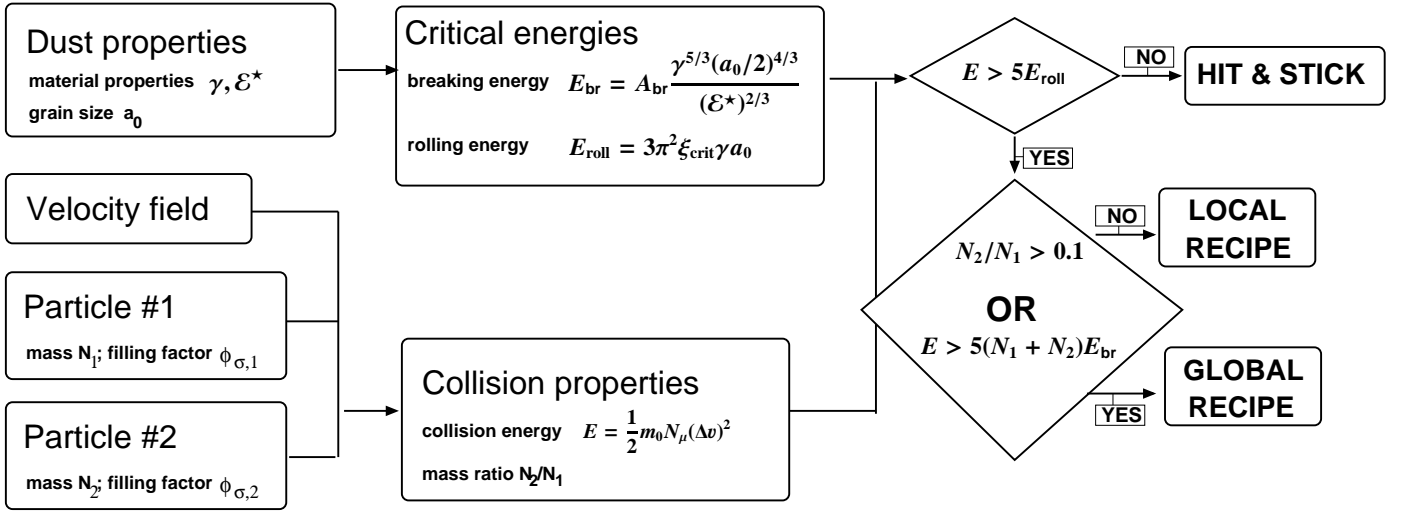}
  \caption{\label{fig:scheme}Schematic decision chain employed to distinguish between the hit-and-stick, global, and local recipes.}
\end{figure*}

Each collisions is classified into one of three groups:
\begin{enumerate}
  \item Hit-and-stick. At low collision energies, the internal structure of the particles is preserved.
  \item Local. Only a small part of the aggregate is affected by the collision, as in, \eg\ erosion. The mass ratio between the two particles is large.
  \item Global. The collision outcome results in a major change to the structure or size of the target aggregate. Relevant for equal-size particles or at large energies.
\end{enumerate}
\Fg{scheme} provides the adopted decision chain between the three regimes. Parameters that enter the chain are the collision energy,
\begin{equation}
  E = \frac{1}{2} m_\mu (\Delta v)^2 = \frac{1}{2}\frac{m_1m_2}{m_1+m_2} (\Delta v)^2,
\end{equation}
where $m_\mu$ the reduced mass, the particle masses ($m_1$ and $m_2$, or, in number of grains, $N_1$ and $N_2$), and the critical energies $E_\mathrm{roll}$ and $E_\mathrm{br}$. Here, $E_\mathrm{br}$ is the energy to break a bond between two grains (the contact) and $E_\mathrm{roll}$ the energy required to roll the contact area over a visible fraction of the grain. These critical energies are defined as \citep{DominikTielens:1997}
\begin{subequations}
  \label{eq:Ebrroll}
\begin{equation}
  E_\mathrm{br} = A_\mathrm{br} \frac{\gamma^{5/3}a_\mu^{4/3}}{\mathcal{E^\star}^{2/3}} \label{eq:Ebr};
\end{equation}
\begin{equation}
  E_\mathrm{roll} = 6\pi^2 \xi_\mathrm{crit} \gamma a_\mu \label{eq:Eroll},
\end{equation}
\end{subequations}
where $a_\mu$ is the reduced radius of the grains ($a_\mu=a_0/2$ for equal size monomers), $\gamma$ the surface energy density of the material, and $\mathcal{E}^\star$ the reduced elastic modulus. Here, following laboratory experiments \citep{BlumWurm:2000} we adopt the values $\xi_\mathrm{crit}=2\times10^{-7}\ \mathrm{cm}$ and $A_\mathrm{br}=2.8\times10^3$, which are larger than the theoretically derived values of \citet{DominikTielens:1997}.  

Thus, when collisional energies are low enough for aggregate restructuring to be unimportant (experimentally determined to be $5E_\mathrm{roll}$; \citealt{BlumWurm:2000}) particles are in the hit-and-stick regime. Similarly, when the collision energy is sufficient to break all contacts the collision falls -- obviously -- in the global regime.  In the remainder the mass ratio of the colliding particles determines whether the collision is global or local. For mass ratios smaller than 0.1 the changes become more localized and it is seen from the simulations that at this point the energy distribution during a collision becomes inhomogeneous. Thus, we take $N_2/N_1=0.1$ as the transition point. A further motivation for adopting this mass ratio is that we construct the global recipe out of simulations between aggregate of the same size. Therefore, the mass range which it represents should not become too large.

Although in our collision model aggregates are characterized by only two properties ($N$ and $\phi_\sigma$), the collision outcome involves many other parameters (discussed in \se{colrec}). These parameters are provided in tabulated form as a function of three input parameters -- dimensionless energy parameter, filling factor, and impact parameter $b$ -- for both the local and the global recipe. To obtain these collision parameters, direct numerical simulations between two colliding aggregates were performed, in which the equations of motions for all grains within the two colliding aggregates are computed \citep{PaszunDominik:2009}.  An example of these quantities is the fraction of missing collisions, which is a result of the fact that we have defined the collision cross section $\sigma^\mathrm{C}$ in terms of the outer radii, $\sigma^\mathrm{C} = \pi (a_\mathrm{out,1}+a_\mathrm{out,2})^2$. 
\app{app2} presents a description of the numerical simulations with their results, discusses a few auxiliary relations that are required for a consistent treatment of the collision model, and treats the format of the collision tables.\footnote{The tables are provided online.}

Two key limitation of these binary aggregate simulations follow from computational constraints: \sumi\ the number of grains that can be included is limited to $N\sim10^3$; and \sumii\ the simulations cannot take into account a grain size distribution that spans over orders of magnitude in mass. These limitations are reflected in our collision model and constitute a potential bottleneck for the level of realism for the application of our results to molecular clouds.  We therefore first motivate our choice for the monomer size and present scaling relations that provide a way to extrapolate the results beyond the parameter space sampled in the simulation.

\subsection{\label{sec:sizea0}Representative monomer size of the MRN grain distribution}
Our recipe is based on simulations of aggregates that are built of monomers of a single size.  Therefore, we treat a monodisperse distribution of grains.  In reality, interstellar dust exhibits a size distribution, or a series of size distributions based on the various grain types \citep[\eg][]{DesertEtal:1990,WeingartnerDraine:2001}.  For simplicity, we compare our monodisperse approach with the MRN grain distribution, in which the number of grains decreases as a $-7/2$ power-law of size, \ie\ $n(a)\mathrm{d}a \propto a^{-7/2}$, between a lower ($a_\mathrm{i}$) and an upper ($a_\mathrm{f}$) size \citep{1977ApJ...217..425M}. Thus, in the MRN-distribution the smallest grains dominate by number, whereas the larger grains dominate the mass.  For an MRN distribution we adopt, $a_\mathrm{i} = 50\ $\AA\ and $a_\mathrm{f} = 0.25\ \micr$. To answer the question which grain size best represents the MRN distribution, we consider both its mechanical and aerodynamic properties.

In the monodisperse situation the mechanical properties of a particle (its strength) can be estimated from the total binding energy per unit mass, $E_\mathrm{br}/m_0$, if we assume each grain has one unique contact.  In the literature the strength of a material is usually denoted by the quantity $Q$. Thus, for a monodisperse model we have
\begin{equation}
  Q_0 = \frac{E_\mathrm{br}(a_0/2)}{m_0} = k \frac{(a_0/2)^{4/3}}{a_0^3} = k 2^{-4/3} a_0^{-5/3},
  \label{eq:Q0}
\end{equation}
where $k=3A_\mathrm{br}\gamma^{5/3}/4\pi\rho_\mathrm{s}{\cal E^\star}^{2/3}$ is a material constant with $\rho_\mathrm{s}$ the bulk density of the material.  A smaller grain size thus lead to significantly stronger aggregates.  For the MRN distribution we assume that a typical contact always involves a small grain, \ie\ $a_\mu \simeq a_i$ enters in the $E_\mathrm{br}$ expression. Assuming again that the number of contacts is of the order of the number of grains, their average strength is given by
\begin{equation}
  Q_\mathrm{MRN} \simeq \frac{E_\mathrm{br}(a_i) \int_{a_i}^{a_f} n(a) \mathrm{d}a}{4\pi\rho_\mathrm{s}/3 \int_{a_i}^{a_f} n(a) a^3 \mathrm{d}a} \simeq k a_\mathrm{i}^{-7/6} a_\mathrm{f}^{-1/2}
  \label{eq:Qmrn}
\end{equation}
where we have used that $a_\mathrm{f} \gg a_\mathrm{i}$.  Equating Eqs.\ (\ref{eq:Q0}) and (\ref{eq:Qmrn}) it follows that the grain size at which the monodisperse model gives aggregates that have the same strength as the MRN is $a_0\simeq2^{4/5} a_i^{7/10} a_f^{3/10} = 560$ \AA.

Apart from the mechanical properties, the \textit{aerodynamic} properties of aggregates are of crucial importance to the collisional evolution. This mainly concerns the initial (fractal) growth stage.  For a single grain $\sigma/m = (3/4\rho_\mathrm{s}) a_0^{-1}$. For the MRN distribution an upper limit on $\sigma/m$ is provided by assuming that all of its surface is exposed, \ie\ as in a 2D arrangement of grains; then,
\begin{equation}
  \frac{\sigma}{m} = \frac{\int_{a_\mathrm{i}}^{a_\mathrm{f}}\pi n(a) a^2 \mathrm{d}a}{\int_{a_\mathrm{i}}^{a_\mathrm{f}}(4\pi\rho_\mathrm{s}/3) n(a) a^3\mathrm{d}a} = \frac{3}{4\rho_\mathrm{s}} (a_i a_f)^{-1/2},
  \label{eq:sofm-mrn}
\end{equation}
and the equivalent aerodynamic grain size becomes $\sqrt{a_\mathrm{i}a_\mathrm{f}} = 350$ \AA. However, this 2D result for the equivalent monodisperse size of the MRN distribution is a considerable underestimation, for three reasons: \sumi\ in 3D the grains will overlap and $\sigma$ becomes lower at the same mass; \sumii\ due to their low rolling energies, the smallest grains of size $a_i$ will already initiate restructure at very low velocities; \sumiii\ in the case of ice-coating, the lower grain size $a_i$ will  be larger by a factor of $\sim$4.

In three dimensions, however, the definition of an equivalent aerodynamic size becomes ambiguous, because $\sigma/m$ is not a constant. To nevertheless get a feeling of the trend, we have calculated the aerodynamic properties of MRN aggregates that consists out of a few big grains, such that their total compact volume is equivalent to a sphere of $0.2-0.3\ \mu\mathrm{m}$.  These MRN aggregates were constructed through a PCA process, \ie\ adding one grain at a time.  Because the aggregates contains the large grains, they fully sample the MRN distribution and can therefore be regarded as the smallest building blocks for the subsequent collisional evolution.

We observed that, due to the above mentioned self-shielding, the aerodynamic size increases to $\sim$$0.08-0.12\ \mu\mathrm{m}$, significantly higher than the 2D limit of \eq{sofm-mrn} \citep[see also][]{Ossenkopf:1993}. Thus, the initial clustering phase of MRN-grains produces structures that behave aerodynamically as compact grains of $\sim$$0.1\ \mu\mathrm{m}$. We remark that this estimate is approximate -- a CCA-like clustering will decrease it, whereas the above mentioned preferential compaction of the very small grains will increase $\sigma/m$ -- but the trend indicates that in 3D the aerodynamic size becomes skewed to the larger grains in the distribution.  Therefore, we take $0.1\ \mu\mathrm{m}$ as the equivalent monomer grain size of the MRN-distribution, but to assess the sensitivity of the adopted grain size to the results we also consider models with a different grain sizes. 

\subsection{\label{sec:par-space-normalization}Scaling of the results}
A key limitation of the aggregate-aggregate collision simulations is the number of grains that can be used; typically, $N\lesssim10^3$ is required in order to complete a thorough parameter study within a reasonable timeframe.  As a consequence, the mass ratio of the colliding aggregate is also restricted.  Furthermore, in the numerical experiments all simulations were performed using material properties applicable to silicates, whereas in molecular clouds we expect the grains to be coated with ice mantles. Clearly, scaling of the results is required such that the findings of the numerical experiments can be applied to aggregates of different size and composition.

Therefore, we scale the collisional energy $E$ to the critical energies, $E_\mathrm{br}$ and $E_\mathrm{roll}$, since these quantities involve the material properties.  For a collision between silicate aggregates and ice-coated aggregates a similar fragmentation behavior may be expected if the collisional energy in the latter case is a factor $E_\mathrm{br}^\mathrm{ice}/E_\mathrm{br}^\mathrm{sil}$ higher. Similarly, restructuring is determined by the rolling energy, $E_\mathrm{roll}$.  Thus, the collision energy is normalized to $E_\mathrm{roll}$ where it concerns the change in filling factor and to $E_\mathrm{br}$ for all other parameters that quantify the collision outcome. 

The division between the global and local recipes is also closely linked to scaling arguments.  In the global recipe energies are normalized to the total number of monomers, $N_\mathrm{tot}$.  Thus, a collision taking place at twice the energy and twice the mass leads to the same fragmentation behavior. However, in the local recipe the amount of damage will be independent of the size of the bigger particle. In this case we then scale by $N_\mu$, essentially the mass of the projectile. This information is captured in a single dimensionless energy parameter $\varepsilon$,
\begin{equation}
  \varepsilon = \frac{E}{N_\mathrm{eff}E_\mathrm{crit}},
  \label{eq:energy-parameter}
\end{equation}
where $E_\mathrm{crit}$ and $N_\mathrm{eff}$ depend on the context: the energy $E_\mathrm{crit}$ can be either one of $E_\mathrm{br}$ or $E_\mathrm{roll}$, whereas $N_\mathrm{eff}$ is one of $N_\mathrm{tot}$ or $N_\mu$ (see \Tb{recipe-quantities}).

\subsection{\label{sec:colrec}The collision parameters}
\label{sec:porting-recipe}

In discussing the collision outcomes, we focus on the local and global recipes, which are a direct result of the numerical simulations. The hit-and-stick recipe is discussed in \app{hitandstick}. To streamline the recipe for a Monte Carlo approach, the specification of the collision outcomes slightly deviates from our previous study \citep{PaszunDominik:2009}.

\begin{figure}
  \includegraphics[width=80mm]{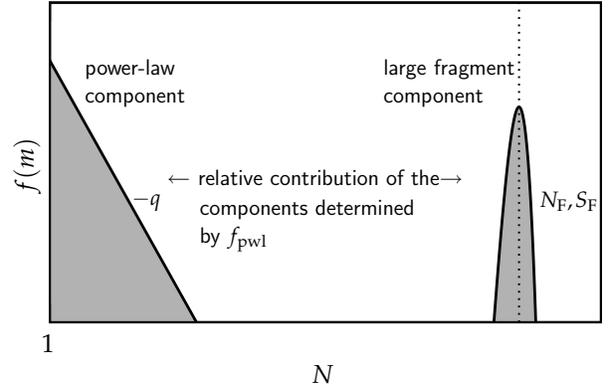}
  \caption{Sketch of the adopted formalism for the size distribution with which the results of the aggregate collision simulation are quantified. See text and \Tb{recipe-quantities} for the description of the symbols.  If $f_\mathrm{pwl}=0$ no power-law component exist. The $N_\mathrm{f}$ and $S_\mathrm{f}$ parameters essentially indicate whether we have zero, one, or two large fragments.}
  \label{fig:quantities-sketch}
\end{figure}
\begin{table*}
  \begin{center}
  \caption{Quantities provided by the adjusted collision recipe.}
  \begin{tabular}{lp{88mm}ll}
    \hline\hline 
    Symbol & Description & \multicolumn{2}{c}{Energy scaling parameters in \eq{energy-parameter}}\\ 
    \cline{3-4} \\[-3.5mm]
           &             & Global       & Local  \\
    (1)    & (2)         & (3)          & (4)    \\
    \hline
    $f_\mathrm{miss}$ & Fraction of collisions that resulted in a miss$^a$        & ---                           & --- \\ 
    $N_\mathrm{f}$    & Mean number of large fragments                            & $N_\mathrm{tot}E_\mathrm{br}$ & $N_\mu E_\mathrm{br}$ \\
    $S_\mathrm{f}$    & Standard deviation of the $N_\mathrm{f}$                  & $N_\mathrm{tot}E_\mathrm{br}$ & $N_\mu E_\mathrm{br}$ \\
    $f_\mathrm{pwl}$  & The fraction of the mass in the small fragments component. Normalized to $N_\mathrm{tot}$ (global recipe) or $N_\mu$ (local recipe). & $N_\mathrm{tot}E_\mathrm{br}$ & $N_\mu E_\mathrm{br}$ \\
    $q$              & Exponent of the power-law distribution of small fragments  & $N_\mathrm{tot}E_\mathrm{br}$ & $N_\mu E_\mathrm{br}$ \\ 
    $C_\phi=\phi_\sigma/\phi_\sigma^\mathrm{ini}$ & Relative change of the geometrical filling factor.& $N_\mathrm{tot}E_\mathrm{roll}$ & $N_\mathrm{tot} E_\mathrm{roll}$ \\
    \hline
  \end{tabular}
  \end{center}
  {\small \textsc{Note}. Columns (3)--(4) denote the energy scaling expressions used to obtain the dimensionless energy parameter, $\varepsilon$, see \eq{energy-parameter}. $^a$ Given as function of $a_\mathrm{out}/a_\sigma$ instead of $\phi_\sigma$, see \app{colltab}.}
  \label{tab:recipe-quantities}
\end{table*}
In the general case of a collision including fragmentation the emergent mass distribution, $f(m)$, consists of two components: \sumi\ a power-law component that describes the small fragments; and \sumii\ a large fragment component that consist out of one or two fragments (see \fg{quantities-sketch}). The separation between the two components is set, somewhat arbitrarily, at a quarter of the total mass of the aggregates, $N_\mathrm{tot} = N_1+N_2$. (It turns out that for our simulations the precise place of the cut is unimportant, because of the lack of severe fragmentation events). Then, the power-law distribution spans the range from monomer mass up to $N=N_\mathrm{tot}/4$, whereas the large-fragment component consists of zero, one, or two aggregates of masses larger than $N_\mathrm{tot}/4$. To obtain the number of large fragments, the recipes provide the mean number of large fragments, $N_\mathrm{f}$, together with its spread $S_\mathrm{f}$. 

\Tb{recipe-quantities} lists all quantities describing a collision outcome. Apart from $N_\mathrm{f}$ and $S_\mathrm{f}$ these include:
\begin{itemize}
  \item The fraction of missing collisions, $f_\mathrm{miss}$. This number gives the fraction of collisions in which no interaction between the aggregates took place. Missing collision are a result from the choice of normalizing the impact parameter $b$ to the outer radius $a_\mathrm{out}$, $\tilde{b} = b/(a_\mathrm{out,1}+a_\mathrm{out,2})$ (see \app{aout-asig}).  For large values of $\tilde{b}$ and very fluffy structures $f_\mathrm{miss}$ becomes significant. 
  \item The mass fraction in the power-law component, $f_\mathrm{pwl}$. It gives the fraction of the total mass ($N_\mathrm{tot}$) that is contained in the power-law component. In the local recipe $f_\mathrm{pwl}$ is defined relative to $N_\mu$, because here the amount of erosion is measured with respect to the smaller projectile.
  \item The exponent of the power-law distribution, $q$. It determines the distribution of the small fragments, \ie\ $f(m) \propto m^{-q}$. 
  \item The relative change in filling factor, $C_\phi$. It gives the change in filling factor of the large fragment component, $\phi_\sigma^\mathrm{new} = C_\phi \phi_\sigma^\mathrm{ini}$. $C_\phi<1$ reflects compaction, whereas $C_\phi>1$ reflects decompaction. Because $C_\phi$ refers to the chance in the filling factor of the large aggregate (for both the local and global recipe), its dimensionless energy parameter $\varepsilon$ is always normalized to the total number of grains, $N_\mathrm{eff} = N_\mathrm{tot}$.  Thus, the compaction may be local and moderate, but the affected quantity -- the filling factor -- describes a global property. Moreover, to prevent possible spuriously high values of $\phi_\sigma$, we artificially assign an upper limit of $33\%$ to the collisional compaction of aggregates \citep{BlumSchraepler:2004}. 
\end{itemize}

\subsubsection{\label{sec:local-rec}The local recipe}
\begin{figure*}[p]
  \centering
  \includegraphics[width=150mm]{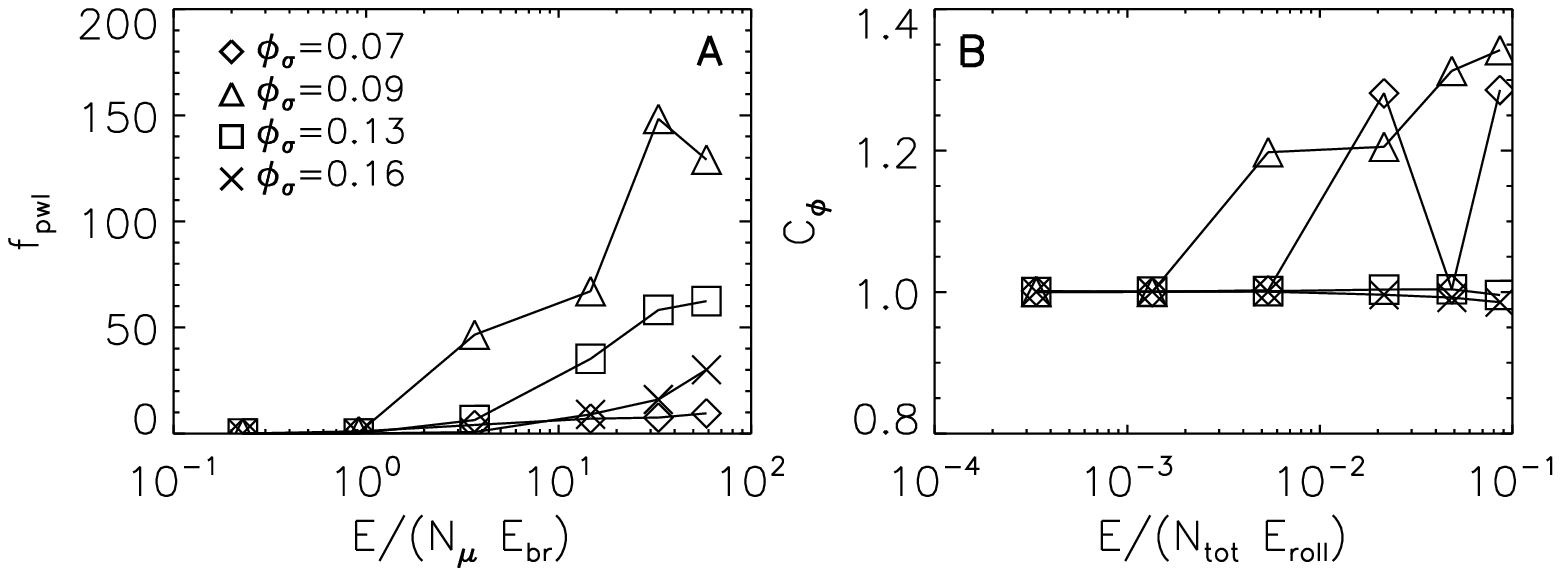}
  \caption{Quantities provided by the local recipe. The left panel shows the mass in small fragments of the power-law component, normalized to the reduced mass of the colliding aggregates $f_\mathrm{pwl} = M_\mathrm{pwl}/m_\mu$.  The right panel shows the relative change in the geometrical filling factor $C_\phi = \phi_\sigma^\mathrm{new}/\phi_\sigma^\mathrm{ini}$. Symbols refer to the initial filling factor of the larger aggregate.}  
  \label{fig:local-recipe}
\end{figure*}
\Fg{local-recipe}a shows how much mass is ejected during collisions at different energies and for different filling factors (symbols). Recall that in the local recipe the $f_\mathrm{pwl}$ quantity involves a normalization to $N_\mu$, rather than $N_\mathrm{tot}$.  At high energies, then, the excavated mass may exceed the mass of the small projectile by even two orders of magnitude.  The distribution of the small fragments created by the erosion is relatively flat with slopes oscillating between $q=-2.0$ and $q=-1.3$.  The number of large fragments $N_\mathrm{F}$ rarely increases above unity. The exception are the `lucky projectiles' that destroy the central contacts of very fluffy aggregates, causing the two sides of the aggregate to become disconnected. If energies are sufficiently high, fragments produced in a cratering event can result in secondary impacts, enhancing the erosion efficiency. 

Since the influence of the impact is local, the change in filling factor is relatively minor (see \fg{local-recipe}b). However, increasing the collision energy results in an increasing degree of compression. Only very fluffy and elongated aggregates may break in half, causing an artificial increase of the filling factor. This can be observed in \fg{local-recipe}b for aggregates with $\phi_\sigma^\mathrm{ini}=0.07$ (diamonds), where the change in filling factor shows a strong variation for energies above $E=10^{-2} N_\mathrm{tot} E_\mathrm{roll}$.

\begin{figure*}[p]
  \centering
  \includegraphics[width=140mm,clip]{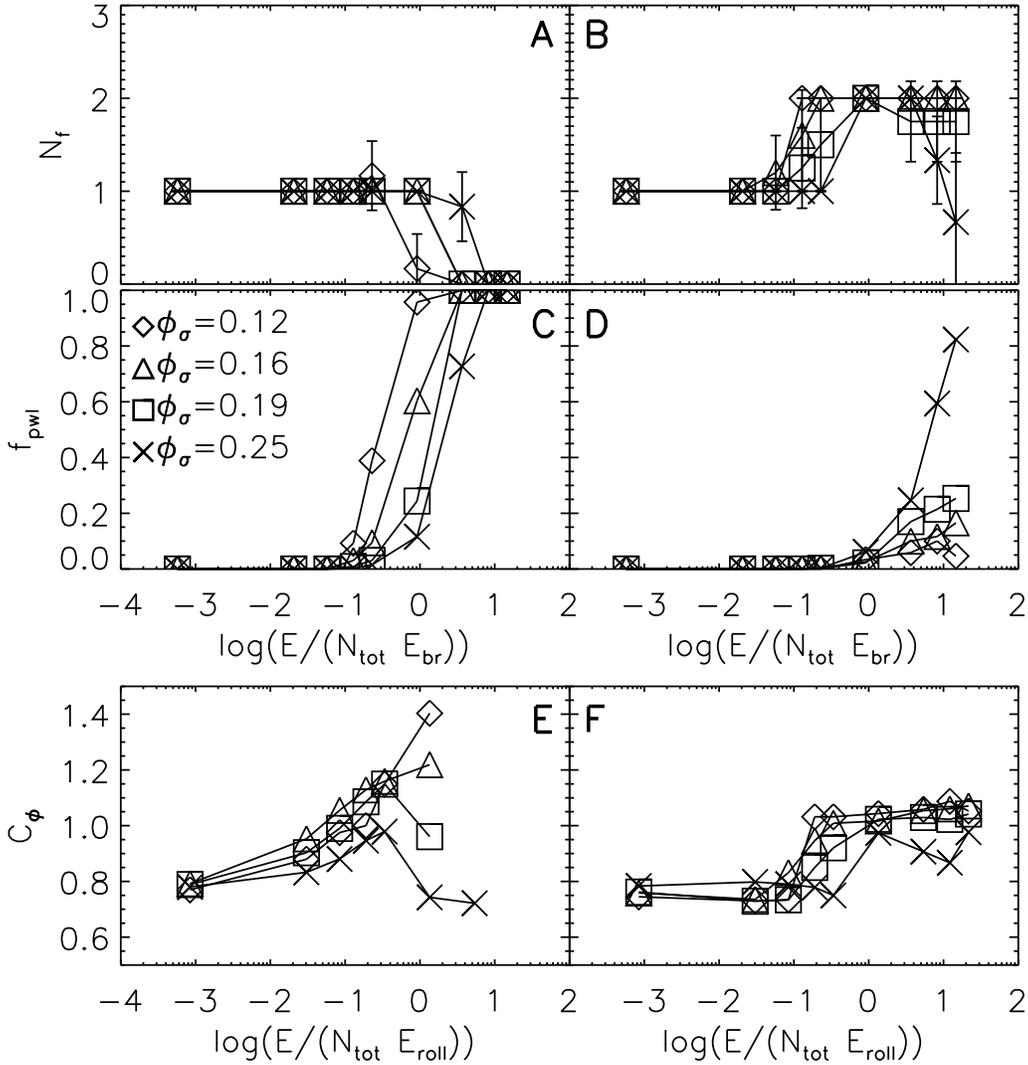}
  \caption{Quantities provided by the global recipe. Left panels correspond to central collisions, while the right panels correspond to off-center collisions at normalized impact parameter $\tilde{b}=0.75$. From top to bottom: number of large fragments $N_\mathrm{f}$ (\textit{A, B}); mass of the small fragments component, $M_\mathrm{pwl}$, normalized to the total mass of the two aggregates $M_\mathrm{tot}$ (\textit{C, D}); relative change in the geometrical filling factor $C_\phi = \phi_\sigma^\mathrm{new}/\phi_\sigma^\mathrm{ini}$ \textit{(E, F)}.}  
  \label{fig:global-recipe}
\end{figure*}
\subsubsection{\label{sec:global-rec}The global recipe}
In \fg{global-recipe} a few results from the global recipe are presented, in which results of collisions at central impact parameter ($\tilde{b}=0$, \textit{left} panels) and off-center collisions ($\tilde{b}=0$, \textit{right} panels) are contrasted.  In Figs.~\ref{fig:global-recipe}a and~\ref{fig:global-recipe}b the number of large particles that remain after a collision, $N_\mathrm{f}$, is given. At low energies the number of fragments is the same ($N_\mathrm{f}=1$) in both cases, reflecting sticking.  At very high energies ($E>5N_\mathrm{tot}E_\mathrm{br}$), the central collision results in catastrophic disruption ($N_\mathrm{f}=0$). Off-center collisions, on the other hand, tend to produce two large fragments at higher energies; because they interact only with their outer parts, the amount of interaction is insufficient to let the colliding aggregates either stick or fragment.

Figures\ \ref{fig:global-recipe}c,d show the mass in the power-law component (small fragments).  Central collisions result in an equal distribution of energy among the monomers. A collision energy of $3 N_\mathrm{tot} E_\mathrm{br}$ is sufficient to shatter an aggregate.  Off-center collisions are more difficult to fully destroy, though, and show, moreover, a strong effect on porosity.  In the most compact aggregate (crosses) over 70\% of the mass ends up in the power-law component, whereas the remainder is in one large fragment. However, these are average quantities, and in some experiments all the mass ended up in the power-law component as can be seen from \fg{global-recipe}b where $N_\mathrm{f}$ drops below unity. For more fluffy aggregates the fragmentation is much less pronounced, because the redistribution of the kinetic energy over the  aggregate is less effective. For example, very fluffy aggregates of filling factor $\phi_\sigma=0.122$ (diamonds) colliding at an impact parameter of $\tilde{b} = 0.75$  produce small fragments which add up to only 6\% of the total mass. The rest of the mass is locked into two large fragments.

The degree of damage can also be assessed through the slope of the power-law distribution of small fragments ($q$, not plotted in \fg{global-recipe}). The steeper the slope, the stronger the damage. Heavy fragmentation produces many small fragments and results in a steepening of the power-law. Although destruction is very strong in the case of a central impact (the slope reaches values of $q = -3.7$ for $E >20 N_\mathrm{tot} E_\mathrm{br}$), it weakens considerably for off-center collisions ($q > -2.0$). For erosive events statistics limit an accurate determination of $q$. However, for erosion the fragments are small in any case, independent of $q$. 

At low energies, the amount of aggregate restructuring, quantified in the $C_\phi$ parameter, is independent of impact parameter (\fg{global-recipe}e,f).  This is simply because the collision energy is too low for restructuring to be significant.  The aggregates' volume then increases in a hit-and-stick fashion, resulting in a decrease of the filling factor ($C_\phi<1$).  With increasing collision energy the degree of restructuring is enhanced, and compression becomes more visible.  Central impacts strongly affect the filling factor $\phi_\sigma$.  \Fg{global-recipe}e shows that the compression is maximal at an impact energy of about $E = N_\mathrm{tot} E_\mathrm{roll}$. Aggregates that are initially compact are difficult to further compress, because for filling factors above a critical value of $33\%$ the required pressures increase steeply \citep{2006ApJ...652.1768B,PaszunDominik:2008}. Any further pressure will preferentially move monomers sideways, causing a flattening of the aggregate and a decreasing packing density.  Off-center collisions, however, lead to a much weaker compression (\fg{global-recipe}f). Here, the forces acting on monomers in the impacting aggregates are more tensile, and tend to produce two large fragments with the filling factor unaffected, $C_\phi=1$.

\section{Results}
\label{sec:results}
\begin{table}
  \caption{\label{tab:pars}List of the model runs. }
  \begin{tabular}{lllll}
    \hline
    \hline
    id            & Density           & Type        & Grain size      & Figure ref.\\
                  & $n$ [cm$^{-3}$]   &             & $a_0$ [$\micr$] &           \\
    (1)           & (2)               & (3)         & (4)             & (5)       \\
    \hline
    1             & $10^3$            & ice               & $0.1  $         &           \\
    2             & $10^4$            & silicates         & $0.1  $         & \fg{pstudy}          \\
    3             & $10^4$            & ice               & $0.1  $         &  \fg{pstudy}         \\
    4             & $10^5$            & silicates         & $0.1  $         & \fg{pstudy}          \\
    5             & $10^5$            & silicates         & $1    $         &           \\
    6$^a$         & $10^5$            & ice               & $0.1  $         & Figs.\ \ref{fig:default}, \ref{fig:defstats}, \ref{fig:por-analys} \\
    7             & $10^5$            & ice               & $1    $         & \fg{a0s}          \\
    8             & $10^5$            & ice               & $0.03 $         & \fg{a0s}          \\
    9             & $10^5$            & ice, compact $^b$ & $0.1  $         & \fg{recip} \\
    10            & $10^5$            & ice, head-on $^c$ & $0.1  $         & Figs.\ \ref{fig:defstats}, \ref{fig:recip}\\
    11            & $10^6$            & silicates         & $0.1  $         & \fg{pstudy}          \\
    12            & $10^6$            & ice               & $0.1  $         & \fg{pstudy}         \\
    13            & $10^7$            & ice               & $0.1  $         & \\
    \hline
  \end{tabular}
  {\small \textsc{Note.} (1) Model number. (2) Number density of the gas. (3) Collision type, describing material parameters and collision setup.  Here, `ice' refers to ice-coated silicates of bulk density identical to bare silicates, $\rho_\mathrm{s} = 2.65\ \mathrm{g\ cm^{-3}}$, but different material properties: $\gamma=370\ \mathrm{erg\ cm^{-2}}$ and $\mathcal{E}^\star=3.7\times10^{10}\ \mathrm{dyn\ cm^{-2}}$. For bare silicates, $\gamma=25\ \mathrm{erg\ cm^{-2}}$ and $\mathcal{E}^\star = 2.8\times10^{11}\ \mathrm{dyn\ cm^{-2}}$. (4) Monomer radius. (5) Figure reference. $^{a}$ The standard model; $^b$ filling factor of particles restricted to a minimum of 33\%; $^c$ central impact collisions only ($b=0$).}
\end{table}

The formulation of the collision recipe in terms of the six output quantities enables us to calculate the collisional evolution by a Monte Carlo method (see \app{mccycle} for its implementation). The sensitivity of the collisional evolution to the environment (\eg\ gas density, grain size, grain type; see \Tb{pars}) is assessed. The coagulation process is generally followed for $10^7\ \mathrm{yr}$. While we realize that bare silicates and the long timescales may not be fully relevant for molecular clouds, we have elected here to extend our calculations to fully probe the characteristics of the coagulation process. In particular, since fragmentation is explicitly included in the collision model we evolve our runs until a steady-state situation materializes.

In \se{anal} the results from the standard model ($n=10^5\ \mathrm{cm^{-3}}$, $a_0=0.1\ \mu\mathrm{m}$, ice-coated silicates) are analyzed. \Se{pstudy} presents the results of our parameter study.

\subsection{\label{sec:anal}The standard model}
\begin{figure}
  \includegraphics[width=\figwidth]{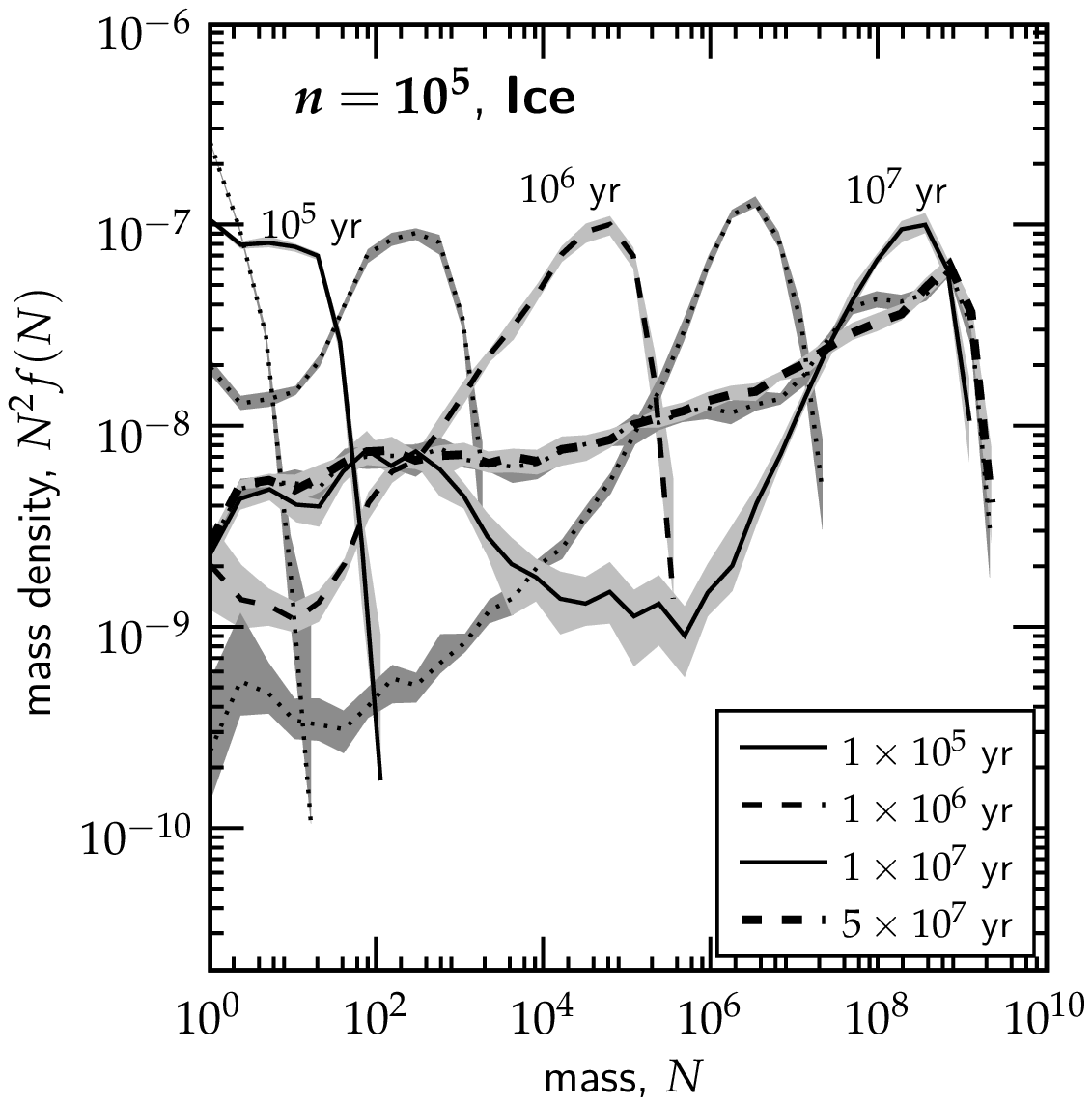}
  \caption{\label{fig:default}Mass distribution of the standard model ($n=10^5\ \mathrm{cm^{-3}}$, $a_0=10^{-5}\ \mathrm{cm}$, ice-coated silicates) at several times during its collisional evolution, until $t=5\times10^7\ \mathrm{yr}$.  The distribution is plotted at times of $10^i\ \mathrm{yr}$ (solid lines, except for the $10^6\ \mathrm{yr}$ curve, which is plotted with a dashed line) and $3\times10^i\ \mathrm{yr}$ (all dotted lines), starting at $t=3\times10^4\ \mathrm{yr}$.  The gray shading denotes the spread in 10 runs.  Mass is given in units of monomers.  The final curve (thick dashed line) corresponds to $5\times10^7\ \mathrm{yr}$ and overlaps the $3\times10^7\ \mathrm{yr}$ curve almost everywhere, indicating that steady-state has been reached.}
\end{figure}
\Fg{default} shows the progression of the collisional evolution of ice-coated silicates at a density of $n=10^5\ \mathrm{cm^{-3}}$ (the standard model) starting from a monodisperse distribution of $0.1\ \mu\mathrm{m}$ grains.  Each curve shows the average of $10$ simulations, where the gray shading denotes the $1\ \sigma$ spread in the simulations.  At $t=0$ the distribution starts out monodisperse at size $N=1$.  The distribution function $f(N)$ gives the number of aggregates per unit volume such that $f(N)\mathrm{d}N$ is the number density of particles in a mass interval $[N,N+\mathrm{d}N]$.  Thus, at $t=0$ the initial distribution has a number density of $f(N=1,t=0) = n \mu m_\mathrm{H}/\mathcal{R}_\mathrm{gd} m_0 = 3.5\times10^{-7}\ \mathrm{cm^{-3}}$ for $n=10^5\ \mathrm{cm^{-3}}$ and $a_0=0.1\ \mu\mathrm{m}$.  On the $y$-axis $N^2 f(N)$ is plotted, which shows the mass of the distribution per logarithmic interval, at several times during the collisional evolution.  The mass where $N^2f(N)$ peaks is denoted the mass peak: it corresponds to the particles in which most of the mass is contained. The peak of the distribution curves stays on roughly the same level during its evolution, reflecting conservation of mass density.

After $10^5\ \mathrm{yr}$ (first solid line) a second mass peak has appeared at $N\simeq10$. The peak at $N=1$ is a result of the compact ($\phi_\sigma=1$) size and smaller collisional cross-section of monomers compared with dimers, trimers.  Furthermore, the high collisional cross section of, \eg\ dimers is somewhat overestimated, being the result of the adopted power-law fit between the geometrical and collisional cross section (\fg{aout-asig}).  These effects are modest, however, and do not affect the result of the subsequent evolution.  Meanwhile, the porosity of the aggregates steadily increases, initially by hit-and-stick collisions but after $\sim$$10^5\ \mathrm{yr}$ mostly through low-energy collisions between equal size particles (global recipe) that do not visibly compress the aggregates. In \fg{por-analys} the porosity distribution is shown at several times during the collisional evolution. Initially, due to low-energy collisions the filling factor decreases as a power-law with exponent $\simeq$$0.3$, $\phi_\sigma \simeq N^{-0.3}$.  This trend ends after $N\sim10^3$, at which time collisions have become sufficiently energetic for compaction to halt the fractal growth. The filling factor then stabilizes and increases only slowly. At $t=3\times10^6\ \mathrm{yr}$ the $N\sim10^7$ particles are still quite porous.

\begin{figure}
  \includegraphics[width=\figwidth]{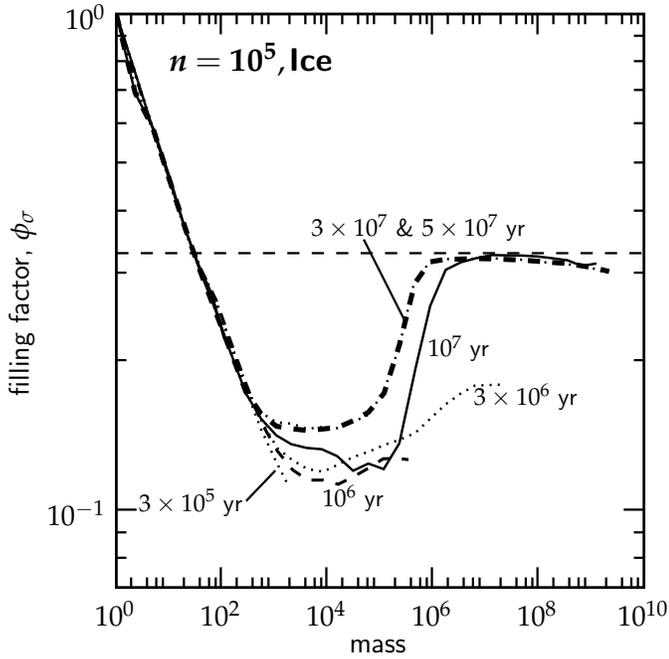}
  \caption{\label{fig:por-analys} The distribution of the filling factor, $\phi_\sigma$, in the standard model, plotted at various times.  Initially, the porosity decreases as a power-law, $\phi \simeq N^{-0.3}$, the fractal regime. Compaction is most severe for the more massive particles where the filling factor reaches the maximum of 33\%. Only mean quantities are shown, not the spread in $\phi_\sigma$.}
\end{figure}
After $3\times10^6\ \mathrm{yr}$ collisions have become sufficiently energetic for particles to start fragmenting, significantly changing the appearance of the distribution (\fg{default}). Slowly, particles at low mass are replenished and growth decelerates. When inquiring the statistics underlying the fragmenting collisions, we find that collisions that result in fragmentation show only modest erosion: only a tiny amount of the mass of the large aggregate is removed.  Therefore, at the onset of erosion, growth is not immediately halted, but it is effective in replenishing the particles at low mass. Eventually, at $N\sim10^9$ ($a\sim100\ \micr$) the erosive collisions reach a point at which there is no net growth.  With increasing time and replenishment, the small particles start to reaccrete to produce a nearly flat distribution in terms of $N^2f(N)$.  Since the final curve ($t=5\times10^7\ \mathrm{yr}$) mostly overlaps the $3\times10^7$ curve (both in \fgs{default}{por-analys}) it follows that steady state is reached on $\sim$$10^7\ \mathrm{yr}$ timescales -- much longer than the timescales on which molecular clouds are thought to exist. 

At $10^7\ \mathrm{yr}$ the largest particles have reached the upper limit of 33\% for the filling factor (see \fg{por-analys}). Compaction increases the collision velocities between the particles and therefore enhances the fragmentation.  The presence of a large population of small particles in the steady state distribution also hints that they are responsible for the higher filling factors particles of intermediate mass (\ie\ $N\sim 10^3-10^6$) have at steady-state compared with the filling factor of these particles at earlier times. Indeed, the turnover point at $N\sim10^3$ corresponds to an energy of $\sim$$5E_\mathrm{roll}$ these particles have with small fragments. Compaction by small particles is thus much more efficient than collisions with larger (but very fluffy) particles.

\subsubsection{Compact and head-on coagulation}
\begin{figure}
  \includegraphics[width=\figwidth]{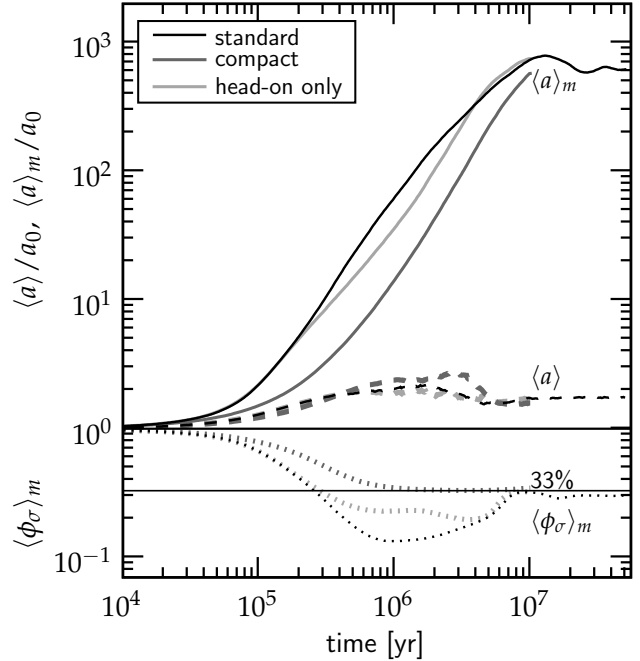}
  \caption{\label{fig:defstats} (\textit{solid} curves) The mean size $\langle a \rangle$ (\textit{dashed} curves), the mass-weighted size $\langle a \rangle_m$ (\textit{dotted} curves) and the mass-weighted filling factor, $\langle \phi_\sigma \rangle_m$ (\textit{solid} curves) of the size distribution as function of time in the standard model (\textit{black} curves), the compact model (\textit{dark gray} curves) and the head-on only model (\textit{light gray} curves).}
\end{figure}
To further understand the influence of the porosity on the collisional evolution, the progression of a few key quantities as function of time are plotted in \fg{defstats}: the mean size $\langle a \rangle$, the mass-average size $\langle a \rangle_m$, and the mass-average filling factor $\langle \phi_\sigma \rangle_m$ of the distribution. Here, mass-average quantities are obtained by weighing the particles of the Monte Carlo program by mass; \eg
\begin{equation}
  \langle a \rangle_m = \frac{\sum_i m_i a_i}{\sum_i m_i},
  \label{eq:mwsize}
\end{equation}
is the mass-weighted size. The weighing by mass has the effect that the massive particles contribute most, because it is usually these particles in which most of the mass resides.  On the other hand, in a regular average all particles contribute equally, meaning that this quantity is particularly affected by the particles that dominate the number distribution. Thus, initially $\langle a \rangle_m = \langle a \rangle$ since at $t=0$ there is only one particle size. With time, however, most of the mass ends up in large particles but the small particles still dominate by number, $\langle a \rangle_m > \langle a \rangle$. This picture is consistent with the distribution plots in \fg{default}.

\begin{figure*}
  \includegraphics[width=\textwidth]{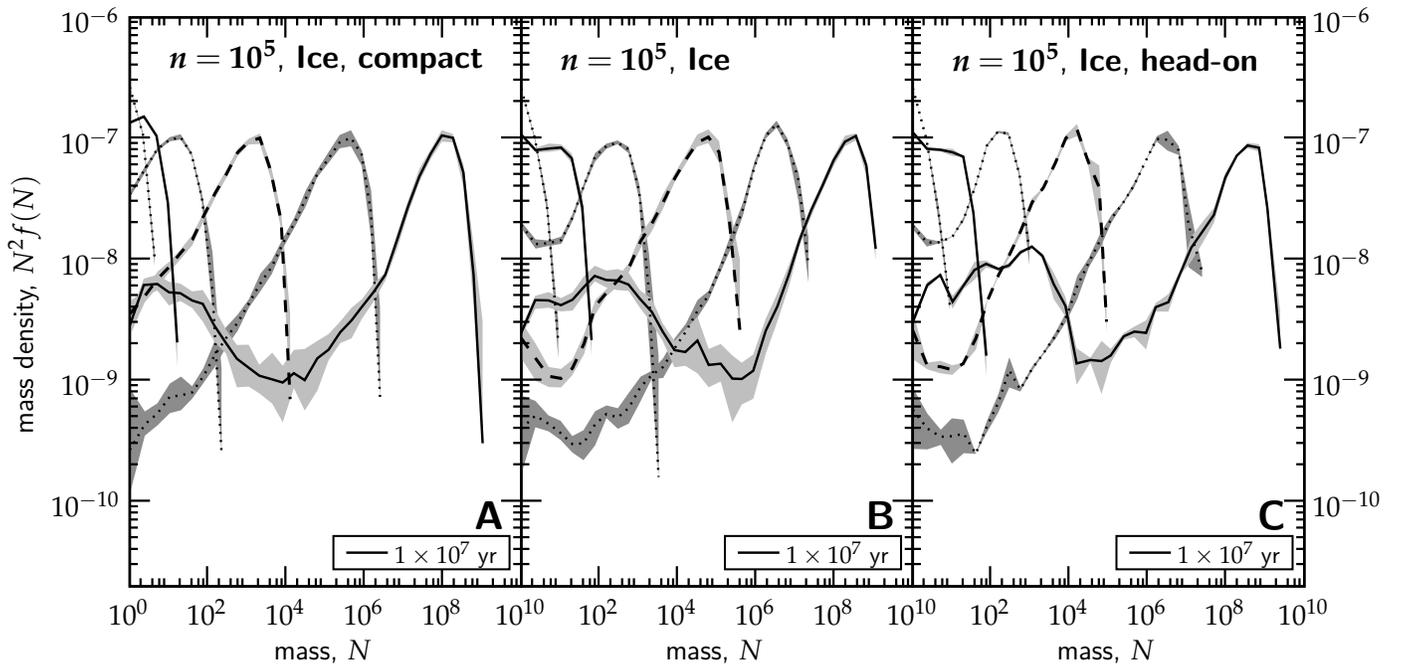}
  \caption{\label{fig:recip}The effects of the collision recipe on the evolution of the size distribution. The standard model (\textit{b}, shown for comparison) is varied and features: (\textit{a}) compact coagulation, in which the filling factor is restricted to a lower limit of 33\%; (\textit{c}) head-on collisions only, where the impact parameter is fixed at $b=0$ for every collision. The calculations last for $10^7\ \mathrm{yr}$.}
\end{figure*}
How sensitive is the emergent size distribution to the adopted collision recipe? To address this question we ran simulations in which the collision recipe is varied with respect to the standard model. The distribution functions of these runs are presented in \fg{recip}, while \fg{defstats} also shows the computed statistical quantities (until $t=10^7\ \mathrm{yr}$). In the case of compact coagulation the filling factor of the particles was restricted to a minimum of 33\% (but small particles like monomers still have a higher filling factor).  Clearly, \fg{defstats} shows that porous aggregates grow during the initial stages (cf.\ also \fg{recip}a and \fg{recip}b). These results are in line with a simple analytic model for the first stages of the growth, presented in \app{simple-anal}: the collision timescales between similar size aggregates is shorter when they are porous. 

\Fg{recip}c presents the results of the standard model in which collisions are restricted to take place head-on, an assumption that is frequently employed in collision studies \citep[\eg][]{WadaEtal:2008,SuyamaEtal:2008}.  That is, except for the missing collision probability ($f_\mathrm{miss}$), the collision parameters are obtained exclusively from the $b=0$ entry.  The temporal evolution of the head-on only model is also given in \fg{defstats} by the light-gray curves.  It can be seen that the particles are less porous than for the standard model.  This follows also from the recipe, see \fg{global-recipe}: at intermediate energies ($E/N_\mathrm{tot} E_\mathrm{roll}\sim 1$) central collisions are much more effective in compacting than off-center collisions.  For the same reason growth in the standard model is also somewhat faster during the early stages.  However, at later times the differences between \fg{recip}b and \fg{recip}c become negligible, indicating that head-on and off-center collisions do not result in a different fragmentation behavior.  

Thus, we conclude that inclusion of porosity significantly boosts the growth rates on molecular cloud relevant timescales ($t=10^5-10^6\ \mathrm{yr}$). Studies that model the growth by compact particles of the same internal density will therefore underestimate the aggregation. Off-center collisions are important to provide a (net) increase in porosity during the restructuring phase but do not play a critical role at later times.

\subsection{\label{sec:pstudy}How density and material properties affect the evolution}
\begin{figure*}
  \centering
  \includegraphics[width=120mm]{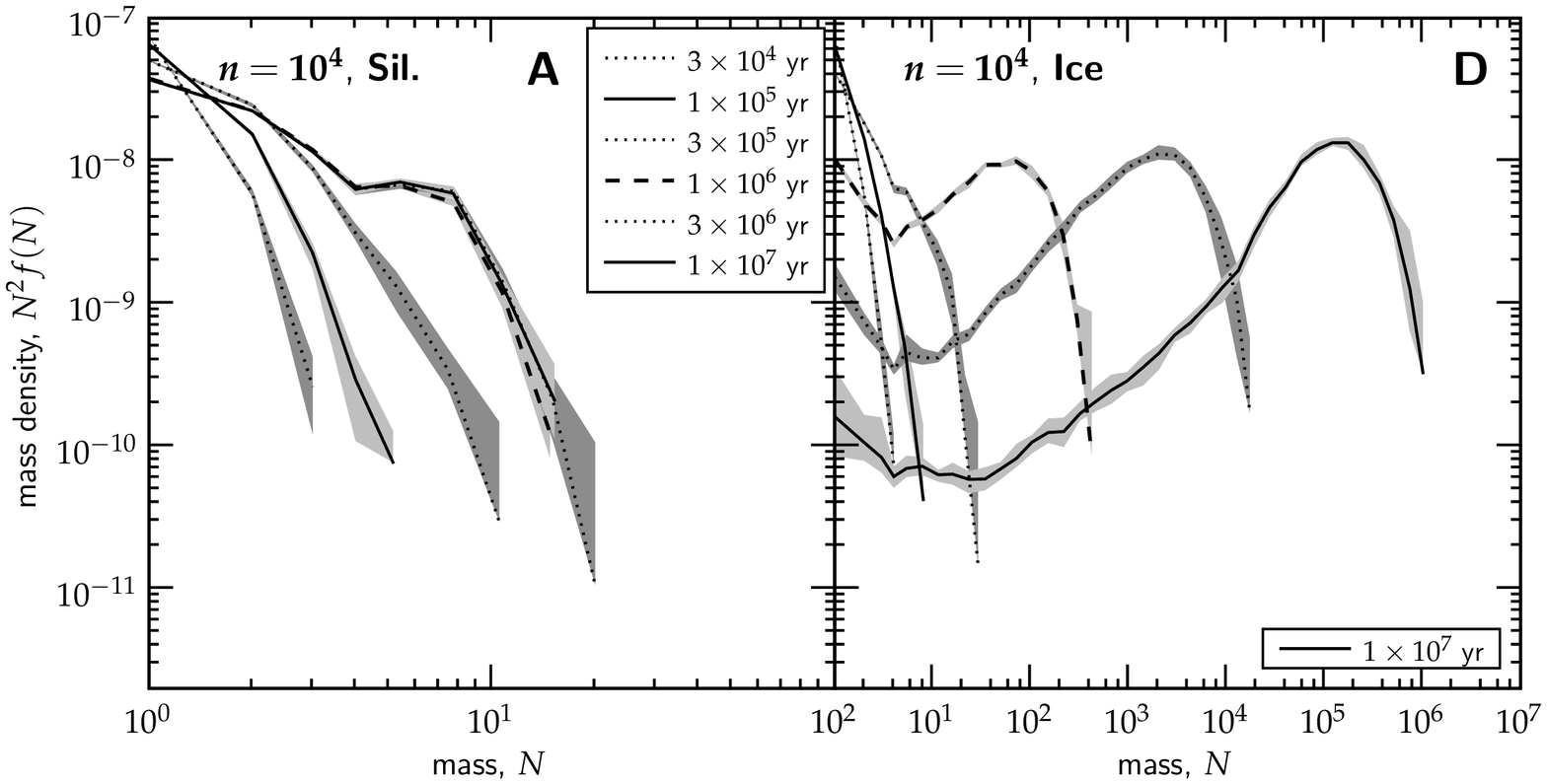}\\
  \includegraphics[width=120mm]{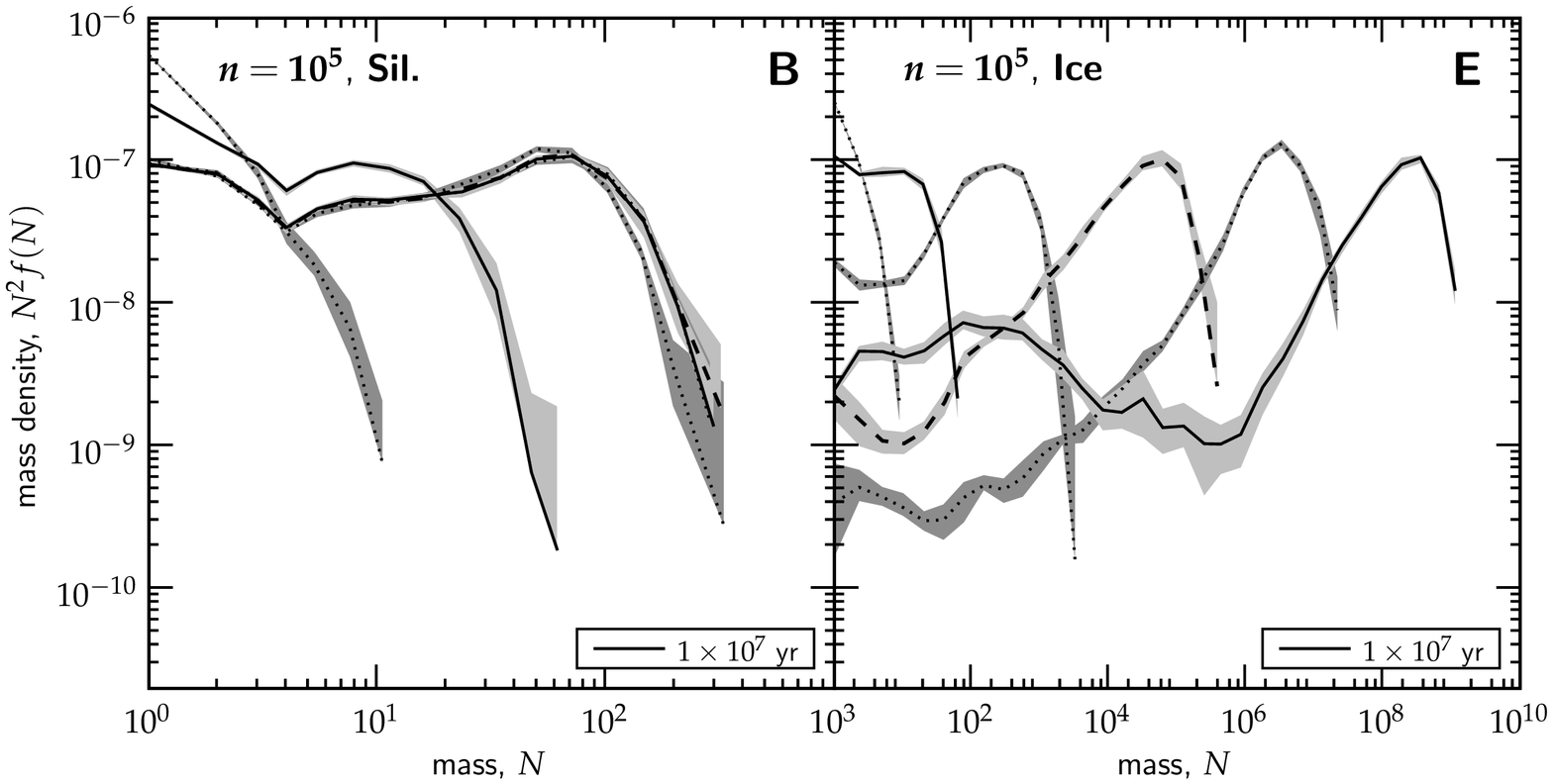}\\
  \includegraphics[width=120mm]{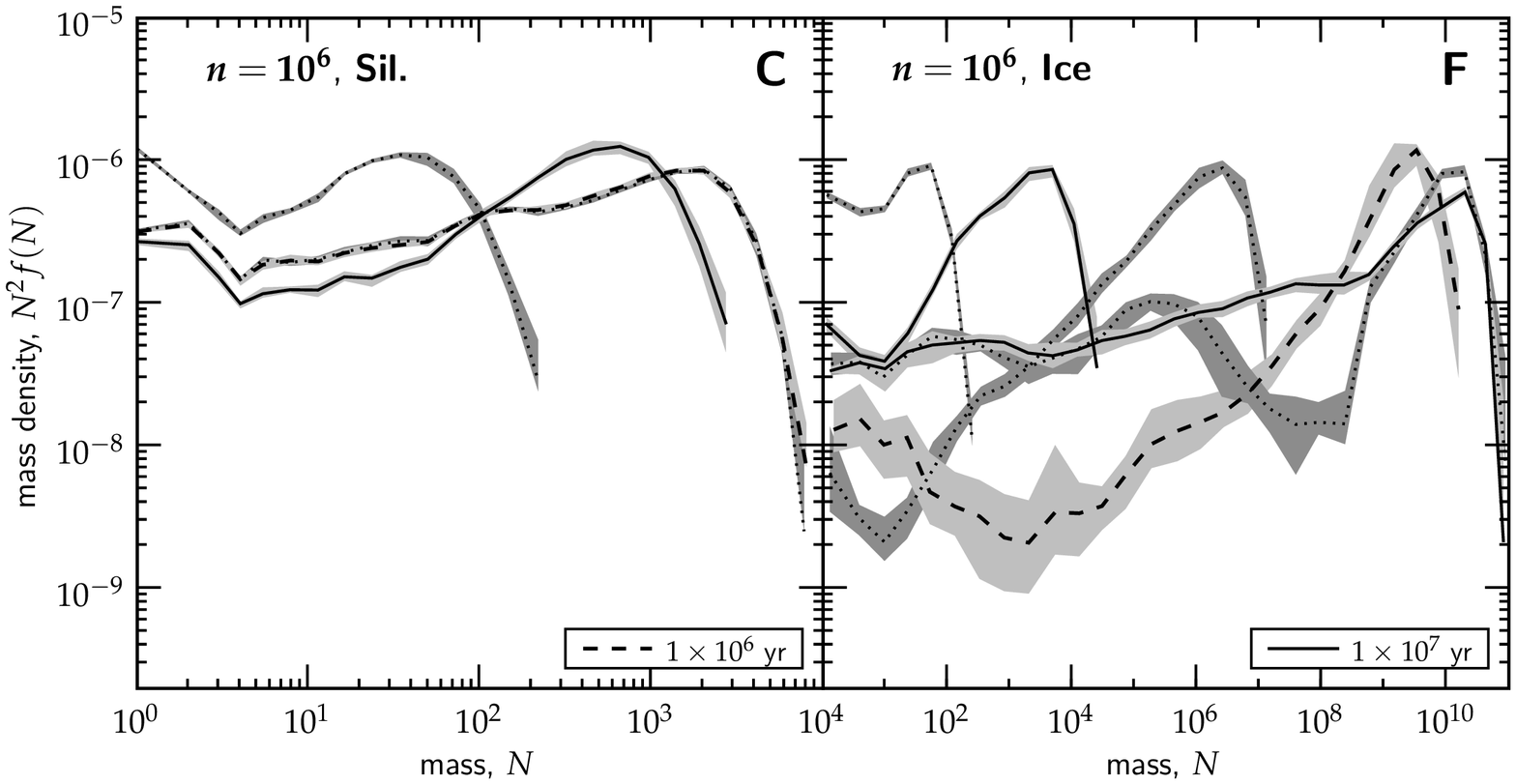}
  \caption{\label{fig:pstudy}Distribution plots corresponding to the collisional evolution of silicates (\textit{left} panels) and ice-coated silicates (\textit{right} panels) at densities of $n=10^4, 10^5$ and $10^6\ \mathrm{cm}^{-3}$ until $t=10^7\ \mathrm{yr}$. For the silicates a steady-state between coagulation and fragmentation is quickly established on timescales of $\sim$$10^6\ \mathrm{yr}$, whereas ice-coated silicates grow much larger before fragmentation kicks in. The initial distribution is monodisperse at $a_0=10^{-5}\ \mathrm{cm}$. Note the different $x$-scaling.}
\end{figure*}
\begin{figure*}[t]
  \includegraphics[width=\textwidth]{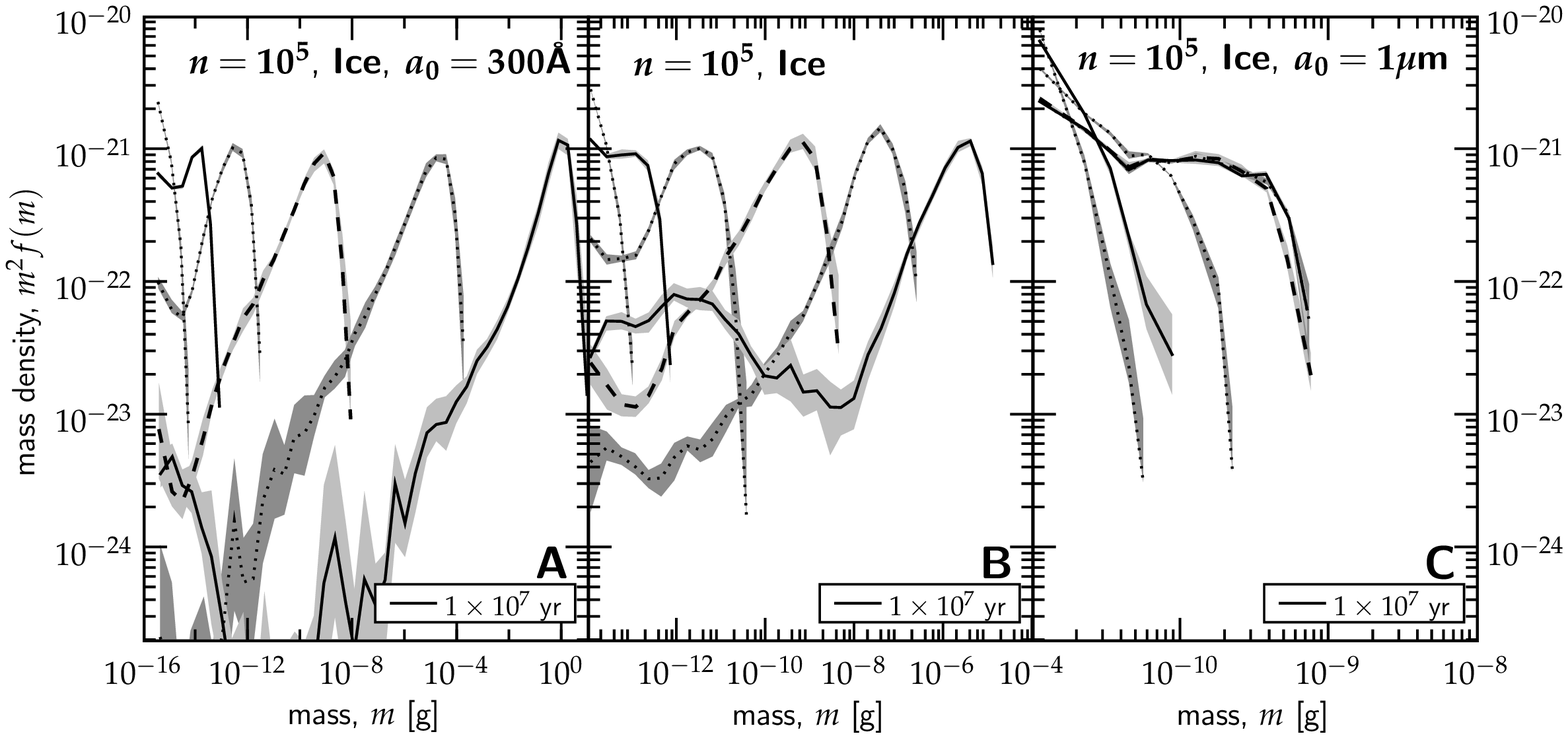}\\
  \caption{\label{fig:a0s}The effects of a different grain size $a_0$ to the collisional evolution: (\textit{a}) $a_0=300\ \mu\mathrm{m}$, (\textit{b}) $a_0=0.1\ \mu\mathrm{m}$ (the default, shown for reasons of comparison), and (\textit{c}) $a_0=1\ \mu\mathrm{m}$. To facilitate the comparison, physical units are used (grams) for the mass of aggregates, rather than the number of monomers ($N$).}
\end{figure*}

\Fg{pstudy}a-c give the collisional evolution of silicates at several densities. In most of the models fragmentation is important from the earliest timescales on. Due to the much lower breaking energy of silicates compared with ice, silicates already start fragmenting at relative velocities of $\sim$10 $\mathrm{m\ s^{-1}}$. As a result, the growth is very modest: only a factor of 10 in size for the $n=10^6\ \mathrm{cm}^{-3}$ model, whereas at lower densities most of the mass stays in monomers. For the same reason, silicates reach steady state already on a timescale of $10^6$ yr, much faster than ice-coated particles.

In the case of ice-coated silicates (\fg{pstudy}d-f) much higher energies are required to restructure and break aggregates and particles grow large indeed. In all cases the qualitative picture reflects that of our standard model (\fg{pstudy}e), discussed in \se{anal}: porous growth in the initial stages, followed by compaction and fragmentation in the form of erosion.  The evolution towards steady-state is a rather prolonged process and is only complete within $10^7\ \mathrm{yr}$ in \fg{pstudy}f. In the low density model of \fg{pstudy}d fragmentation does not occur within $10^7\ \mathrm{yr}$. Steady state is characterized by a rather flat mass spectrum.  

In \fg{a0s} the collisional evolution is contrasted for three different monomer sizes: $a_0=300$\ \AA\ (\fg{a0s}a), $0.1\ \mu\mathrm{m}$ (the standard model, \fg{a0s}b), and 1\ $\mu\mathrm{m}$ (\fg{a0s}c).  To obtain a proper comparison, \fg{a0s} uses physical units (grams) for the mass of the aggregates, rather than the dimensionless number of monomers, $N$. From \fg{a0s} it can be seen that the models are extremely sensitive to the grain size.  In \fg{a0s}c, for example, the weaker aggregates result in the dominance of fragmenting collisions already from the start.  These curves, therefore, resemble the silicate models of \fg{pstudy}b.  

\Fg{a0s}a, on the other hand, shows that a reduction of the grain size by about a factor three ($a_0=0.03\ \mu\mathrm{m}$) enhances the growth significantly. Despite starting from a lower mass, the 300 \AA\ model overtakes the standard model at $t\sim10^6\ \mathrm{yr}$. An understanding of this behavior is provided in \app{simple-anal}, the key element being the persistence of the hit-and-stick regime from which it is very difficult to break out if $a_0$ is small. Until $4\times10^6\ \mathrm{yr}$  visible compaction fails to take place and aggregates become very porous indeed ($\phi\simeq4\times10^{-4}$). The consequence is that fragmentation is also delayed, and has only tentatively started near the end of the simulations.  We caution, however, against the relevance of the 300 \AA\ model; as explained in \se{sizea0}, the choice of $a_0=300$ \AA\ is too low to model aerodynamic and mechanical properties of MRN aggregates. But \fg{a0s} serves the purpose of showing the sensitivity of the collisional result on the underlying grain properties.

\subsection{Comparison to expected molecular cloud lifetimes}
\begin{table*}
  \begin{center}
  \caption{\label{tab:growth-table} Mass-weighted size of the distribution, $\langle a \rangle_m$, at several distinct events during the simulation run. }
\begin{tabular}{lllllll}
  \hline
  \hline
  & \multicolumn{6}{c}{$\langle a \rangle_m$ [cm]} \\
  \cline{2-7} \\[-3.5mm]
  model &  $10^4\ \mathrm{yr}$&  $10^5\ \mathrm{yr}$&  $10^6\ \mathrm{yr}$&  $10^7\ \mathrm{yr}$ & $t_\mathrm{ff}(n)$ & $t_\mathrm{ad}(n)$ \\
  (1) & (2) & (3) & (4) & (5) & (6) & (7)\\
  \hline
$n=10^3$, ice & {$1.0(-5)$} & {$1.0(-5)$} & {$1.2(-5)$} & {$8.3(-5)$} & {$1.2(-5)$} & {$8.3(-5)$} \\ 
$n=10^4$, silicates & {$1.0(-5)$} & {$1.1(-5)$} & {$1.4(-5)$} & {$1.4(-5)$} & {$1.2(-5)$} & {$1.4(-5)$} \\ 
$n=10^4$, ice & {$1.0(-5)$} & {$1.1(-5)$} & {$4.6(-5)$} & {$8.5(-4)$} & {$1.5(-5)$} & {$8.5(-4)$} \\ 
$n=10^5$, silicates & {$1.0(-5)$} & {$1.9(-5)$} & {$4.0(-5)$} & {$4.0(-5)$} & {$2.0(-5)$} & {$4.0(-5)$} \\ 
$n=10^5$, silicates, $a_0=10^{-4}$ & {$1.0(-4)$} & {$1.0(-4)$} & {$1.0(-4)$} & {$1.0(-4)$} & {$1.0(-4)$} & {$1.0(-4)$} \\ 
$n=10^5$, ice & {$1.0(-5)$} & {$2.2(-5)$} & {$6.4(-4)$} & {$7.4(-3)$} & {$2.3(-5)$} & {$3.2(-3)$} \\ 
$n=10^5$, ice, $a_0=10^{-4}$ & {$1.0(-4)$} & {$1.1(-4)$} & {$2.2(-4)$} & {$2.3(-4)$} & {$1.1(-4)$} & {$2.3(-4)$} \\ 
$n=10^5$, ice, $a_0=3\times10^{-6}$ & {$3.2(-6)$} & {$1.1(-5)$} & {$1.3(-3)$} & {$4.3$} & {$1.2(-5)$} & {$2.4(-1)$} \\ 
$n=10^5$, ice, compact & {$1.0(-5)$} & {$1.5(-5)$} & {$1.4(-4)$} & {$5.8(-3)$} & {$1.6(-5)$} & {$1.3(-3)$} \\ 
$n=10^5$, ice, head-on & {$1.0(-5)$} & {$2.2(-5)$} & {$3.6(-4)$} & {$7.5(-3)$} & {$2.4(-5)$} & {$3.1(-3)$} \\ 
$n=10^6$, silicates & {$1.4(-5)$} & {$1.2(-4)$} & {$1.3(-4)$} & {$1.3(-4)$} & {$4.4(-5)$} & {$1.3(-4)$} \\ 
$n=10^6$, ice & {$1.4(-5)$} & {$2.7(-4)$} & {$3.7(-2)$} & {$2.0(-2)$} & {$4.6(-5)$} & {$2.9(-2)$} \\ 
$n=10^7$, ice & {$7.9(-5)$} & {$3.7(-2)$} & {$5.2(-2)$} & {$6.1(-2)$} & {$8.6(-5)$} & {$7.8(-1)$} \\ 
  \hline
\end{tabular}
  \end{center}
  {\small \textsc{Note.} Column\ (1) lists the models in terms of the density ($n$) and material properties. The monomer size ($a_0$) is $0.1\ \micr$, unless otherwise indicated. Cols.\ (2)--(5) give the mass-weighted size of the distribution at fixed coagulation times. Likewise, cols.\ (6)--(7) provide $\langle a \rangle_m$ at the free-fall and the ambipolar diffusion timescale of the cloud that corresponds to the gas density $n$. These are a function of density and are given in \eq{tff} and \eq{tad}, respectively. Values $a\times10^b$ are denoted $a(b)$.}
\end{table*}
\begin{table*}
  \centering
  \caption{\label{tab:kappa-table}Like \Tb{growth-table} but for the geometrical opacity $\kappa$ of the particles.}
\begin{tabular}{lllllll}
  \hline
  \hline
  & \multicolumn{6}{c}{$\langle \kappa \rangle\ [\mathrm{cm}^2\ \mathrm{g}^{-1}]$} \\
  \cline{2-7} \\[-3.5mm]
  model &  $10^4\ \mathrm{yr}$&  $10^5\ \mathrm{yr}$&  $10^6\ \mathrm{yr}$&  $10^7\ \mathrm{yr}$  & $t_\mathrm{ff}(n)$ & $t_\mathrm{ad}(n)$\\
  (1) & (2) & (3) & (4) & (5) & (6) & (7)\\
  \hline
$n=10^3$, ice & {$2.8(4)$} & {$2.8(4)$} & {$2.7(4)$} & {$1.5(4)$} & {$2.7(4)$} & {$1.5(4)$} \\ 
$n=10^4$, silicates & {$2.8(4)$} & {$2.8(4)$} & {$2.7(4)$} & {$2.6(4)$} & {$2.7(4)$} & {$2.6(4)$} \\ 
$n=10^4$, ice & {$2.8(4)$} & {$2.8(4)$} & {$2.0(4)$} & {$2.5(3)$} & {$2.6(4)$} & {$2.5(3)$} \\ 
$n=10^5$, silicates & {$2.8(4)$} & {$2.5(4)$} & {$2.0(4)$} & {$2.0(4)$} & {$2.5(4)$} & {$2.0(4)$} \\ 
$n=10^5$, silicates, $a_0=10^{-4}$ & {$2.8(3)$} & {$2.8(3)$} & {$2.8(3)$} & {$2.8(3)$} & {$2.8(3)$} & {$2.8(3)$} \\ 
$n=10^5$, ice & {$2.8(4)$} & {$2.4(4)$} & {$5.1(3)$} & {$2.3(3)$} & {$2.4(4)$} & {$8.4(2)$} \\ 
$n=10^5$, ice, $a_0=10^{-4}$ & {$2.8(3)$} & {$2.8(3)$} & {$2.4(3)$} & {$2.4(3)$} & {$2.8(3)$} & {$2.4(3)$} \\ 
$n=10^5$, ice, $a_0=3\times10^{-6}$ & {$9.3(4)$} & {$7.1(4)$} & {$1.4(4)$} & {$4.4(2)$} & {$6.9(4)$} & {$1.7(3)$} \\ 
$n=10^5$, ice, compact & {$2.8(4)$} & {$2.6(4)$} & {$8.0(3)$} & {$1.9(3)$} & {$2.6(4)$} & {$1.0(3)$} \\ 
$n=10^5$, ice, head-on & {$2.8(4)$} & {$2.4(4)$} & {$4.9(3)$} & {$3.1(3)$} & {$2.4(4)$} & {$9.3(2)$} \\ 
$n=10^6$, silicates & {$2.7(4)$} & {$1.4(4)$} & {$1.4(4)$} & {$1.4(4)$} & {$2.0(4)$} & {$1.4(4)$} \\ 
$n=10^6$, ice & {$2.7(4)$} & {$1.2(4)$} & {$6.7(2)$} & {$2.2(3)$} & {$2.0(4)$} & {$1.5(3)$} \\ 
$n=10^7$, ice & {$1.7(4)$} & {$1.8(3)$} & {$1.4(3)$} & {$1.7(3)$} & {$1.6(4)$} & {$6.4(2)$} \\ 
  \hline
\end{tabular}
\end{table*}
Tables\ \ref{tab:growth-table} and \ref{tab:kappa-table} present the results of the collisional evolution in tabular format. In \Tb{growth-table} the mass-weighted size of the distribution ($\langle a \rangle_m$, reflecting the largest particles) is given, and in \Tb{kappa-table} the opacity of the distribution is provided, which reflects the behavior of the small particles. Here, opacity denotes geometrical opacity -- the amount of surface area per unit mass -- which would be applicable for visible or UV radiation, but not to the IR. Its definition is, accordingly,
\begin{equation}
  \langle \kappa \rangle = \frac{\sum_i \pi a_{\sigma,i}^2}{\sum_i m_i},
\label{eq:kappa}
\end{equation}
where the summation is over all particles in the simulation. These tables show, for example, that in order to grow chondrule-size particles ($\sim$$10^{-3}\ \mathrm{g}$), dust grains need to be ice-coated and, except for the $n=10^6\ \mathrm{cm^{-3}}$ model, coagulation times of $\sim$$10^7\ \mathrm{yr}$ are needed. 

To further assess the impact of grain coagulation we must compare the coagulation timescales to the lifetimes of molecular clouds.  In a study of molecular clouds in the solar neighborhood \citet{2001ApJ...562..852H} hint that the lifetime of molecular cloud is short, because of two key observational constraints: \sumi\ most cores do contain young stars, rather than being starless; and \sumii\ the age of the young stars that are still embedded in a cloud is $1-2$ Myr at most. From these two arguments it follows that the duration of the preceding starless phase is also $1-2$ Myr. If core lifetimes are limited to the free-fall time (\eq{tff}), then, the grain population will not leave significant imprints on either \sumi\ the large particles produced, or \sumii\ the removal of small particles. This can be seen from Tables\ \ref{tab:growth-table} and \ref{tab:kappa-table} where $\langle a \rangle_m$ and $\langle \kappa \rangle$ are also given at the free-fall time of the simulation (col.\ 6). From \Tb{growth-table} it is seen that the sizes of the largest particles all stay below $1\ \micr$ (except the models that started already with a monomer sizes of $a_0=1\ \micr$). Likewise, \Tb{kappa-table} shows that the opacities from the $t_\mathrm{ff}$ entry are similar to those of the `initial' $10^4\ \mathrm{yr}$ column, \ie\ growth is negligible on free-fall timescales.

\begin{figure}
  \centering
  \includegraphics[width=88mm]{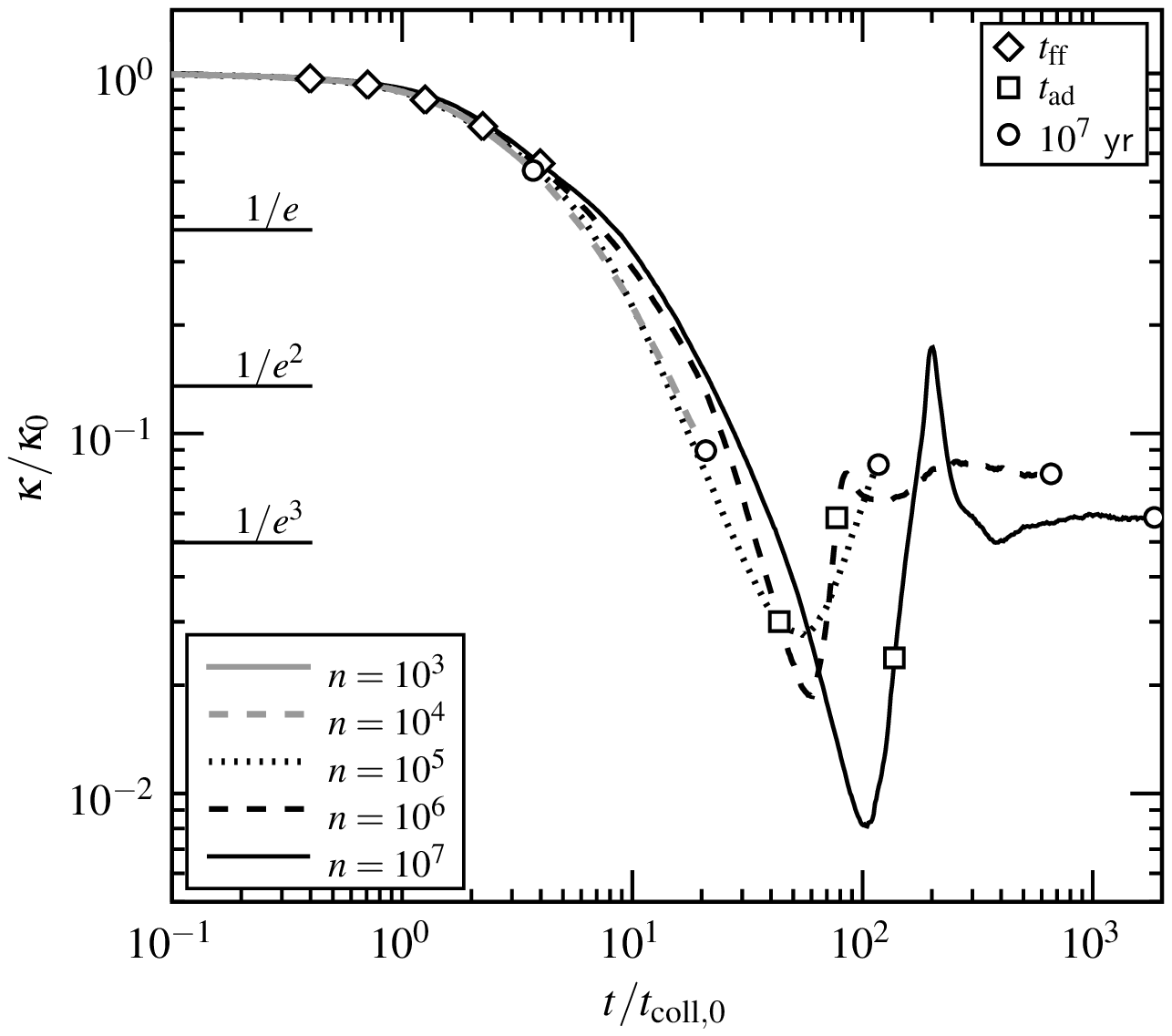}
  \caption{\label{fig:tscales}The opacity $\kappa$ normalized to its initial value \vs\ time in units of the initial collision time $t_\mathrm{coll,0}$ (\eq{tcoll}) for the ice-coated, $a_0=0.1\ \micr$ silicates models at five different gas densities $n$. The decrease in opacity occurs on timescales of $\sim$$10t_\mathrm{coll,0}$. In simulations where small grains reemerge due to fragmentation $\kappa$ starts to increase again. The free-fall (\textit{diamonds}) and ambipolar diffusion timescales (\textit{squares}) are indicated as far as these fall within $10^7$ yr (\textit{circles}). Points of low density appear at lower $t/t_\mathrm{coll,0}$.}
\end{figure}
This information is also displayed in \fg{tscales}, where the opacity with respect to the initial opacity, $\kappa/\kappa_0$, is plotted against time for all densities from the $a_0=10^{-5}\ \mathrm{cm}$ ice-coated silicate models. In \fg{tscales} time is normalized to the initial collision timescale between two grains, $t_\mathrm{coll,0}$, which is a function of density (see \eq{tcoll}).  The similarity of the curves for the first $\sim$10 $t_\mathrm{coll,0}$ is in good agreement with a simple analytic model presented in \app{simple-anal}. In models where small particles are replenished by fragmentation, $\kappa$ first obtains a minimum and later levels-off at $\kappa/\kappa_0\sim0.05$. Also in \fg{tscales}, the free-fall and ambipolar diffusion timescales are indicated with diamond and square symbols, respectively. Due to the normalization by $t_\mathrm{coll,0}$ these occur within a relatively narrow region, despite the large range in densities considered. It is then clear that at a free-fall timescale no significant reduction of the opactiy takes place, since $t_\mathrm{ff}/t_\mathrm{coll,0} \lesssim 1$. 

However, there is still a lively debate whether the fast SF picture -- or, rather, a short lifetime for molecular clouds -- is generally attainable, as cores may have additional support mechanisms \citep{2004ApJ...616..283T}.  If clouds are magnetically supported, the collapse is retarded by ambipolar diffusion (AD),  and the relevant timescales are much longer than the free-fall timescale (see \eq{tad}), $t_\mathrm{AD}/t_\mathrm{coll,0} \gg1$ (\fg{tscales}) . Then, growth becomes significant, as can be seen from \Tb{growth-table} where aggregrates reach sizes of $\sim$100 $\micr$ in the densest models on an AD-timescale. For the highest density models timescales are even sufficiently long for fragmentation to replenish the small grains. (Note that, although $t_\mathrm{AD}$ decreases with density, the evolution of the core is determined by the quantity $t_\mathrm{AD}/t_\mathrm{coll,0}$, which increases with $n$.) Thus, if cores evolve on AD-timescales, the observational appearance of the core will be significantly affected.  \Tb{kappa-table} and \fg{tscales} show that the  UV-opacity, which is directly proportional to $\kappa$, will be reduced by a factor of $\sim$10. Studies that relate the $A_\mathrm{V}$ extinction measurements to column densities through the standard dust-to-gas ratio therefore could underestimate the amount of gas that is actually present. 

\section{Discussion}
\label{sec:discus}
\subsection{Growth characteristics and comparison to previous works}
In his pioneering work to the study of dust coagulation in molecular clouds, \citet{Ossenkopf:1993}, like our study, follows the internal structure of particles and presents a model for the change in particle properties for collisions in the hit-and-stick regime.  Furthermore, the grains are characterized by an MRN size distribution. The model of \citet{Ossenkopf:1993} only treats the hit-and-stick collision regime but at the high densities ($n_\mathrm{H}\ge10^6\ \mathrm{cm^{-3}}$) and short timescales ($\sim$$10^5\ \mathrm{yr}$) he considers any compaction or fragmentation between ice(-coated) particles is indeed of no concern.  The coagulation then proceeds to produce particles of compact size $\sim$$0.5\ \mu\mathrm{m}$ at $n_\mathrm{H}=10^6\ \mathrm{cm}^{-3}$.  It can be seen from \Tb{growth-table} that the growth in the corresponding model of our study (ice, $n=10^6\ \mathrm{cm}^{-3}$) is higher: $2.7\ \mu\mathrm{m}$. This large difference (especially in terms of mass) can be attributed to the fact that \citet{Ossenkopf:1993} ignores turbulent relative velocities between particles of friction times $\tau_\mathrm{f}<t_\mathrm{s}$. As a result, growth is predominantly PCA, because the small grains can only be swept up by bigger aggregates, rendering his coagulation more compact in comparison to our model and therefore slower.  Additionally, due to the different definitions we use for `density' ($n$ \vs\ $n_\mathrm{H}$, see footnote \ref{foot:gasdensity}) our `$10^6\ \mathrm{cm}^{-3}$ model' is denser by a factor of 1.7, resulting in a lower collision timescale and faster growth.

However, at timescales $t>t_\mathrm{coll,0}$ (where $t_\mathrm{coll,0}$ for a distribution would be the collision time between big grains) hit-and-stick growth will turn into CCA. Consequently, fast growth is expected on timescales larger than a collision timescale (see \app{app1}). By $10^5$ yr this condition has clearly been fulfilled in our $n=10^6\ \mathrm{cm}^{-3}$ model, but it is likely that, due to the above mentioned differences, it has not been met, or perhaps only marginally, in \citet{Ossenkopf:1993}. Thus, rather than fixing on one point in time, a more useful comparison would be to compare the growth curves, $a(t)$.

On the other hand, \citet{WeidenschillingRuzmaikina:1994}, adopt a Bonnor-Ebert sphere to model the molecular cloud, and calculate the size distribution for much longer timescales ($t=10^7\ \mathrm{yr}$). Like our study, \citet{WeidenschillingRuzmaikina:1994} include fragmentation in the form of erosion and, at high energies, shattering.  Their particles are characterized by a strength of $Q\sim10^6\ \mathrm{erg\ g^{-1}}$, which are, therefore, somewhat weaker than the particles of our standard model. Although their work lacks a dynamic model for the porosity evolution, it is assumed that the initial growth follows a fractal law until $30\ \micr$. At these sizes the minimum filling factor becomes less than 1\%, lower than our results.  On timescales of $\sim$$10^7\ \mathrm{yr}$ particles grow to $\gtrsim$$100\ \micr$, comparable to that of our standard model. 

A major difference between \citet{WeidenschillingRuzmaikina:1994} and our works concerns the shape of the size distribution. Whereas in our calculations the mass-peak always occurs at the high-mass end of the spectrum, in the \citet{WeidenschillingRuzmaikina:1994} models most of the mass stays in the smallest particles.  Perhaps, the lack of massive particles in the \citet{WeidenschillingRuzmaikina:1994} models is the result of the spatial diffusion processes this work includes; massive particles, produced at high density, mix with less massive particles from the outer regions.  In contrast, our findings regarding steady-state distributions agree qualitatively with the findings of \citet{2008A&A...480..859B} for protoplanetary disks.  Despite the different environments, and therefore different velocity field, we find that the steady state coagulation-fragmentation mass spectrum is characterized by a rather flat $m^2f(m)$ mass function. 

It is also worthwhile to compare the aggregation results from our study with the constituent particles of meteorites, chondrules ($a\sim300\ \mu\mathrm{m}$) and calcium-aluminium inclusions (CAIs, $a\sim \mathrm{cm}$). Although most meteoriticists accept a nebular origin for these species \citep[\eg][]{HussEtal:2001}, \citet{Wood1998} suggested that, in order to explain Al-26 \textit{free} inclusions, aggregates the sizes of CAIs (and therefore also chondrules), formed in the protostellar cloud. These large aggregates then were self-shielded from the effects of the Al-26 injection event. However, our results indicate that growth to cm-sizes seems unlikely. Only the dense ($n\ge10^6$) models can produce chondrule-size progenitors and only at a (long) ambipolar diffusion timescale.

\subsection{Observational implications for molecular clouds}
In our models we observe that the shape of the initially monodisperse dust size distribution evolves first to a Gaussian-like distribution and eventually to a flat steady-state distribution.  For timescales longer than the coagulation timescale (\eq{tcoll}) we can expect that this result is independent of the initial conditions, even if the coagulation starts from a power-law distribution.   Essentially, these distributions are a direct result of the physics of the coagulation: the Gaussian-like distribution reflects the hit-and-stick nature of the agglomeration process at low velocities while the `flat' $N^2f(N)$ distribution at later times results from a balance between fragmentation -- erosion but not catastrophic destruction -- and growth.  In contrast, in interstellar shocks grains acquire much larger relative velocities and grain-grain collisions will then quickly shatter aggregates into their constituent monomers \citep{JonesEtal:1996,HirashitaYan:2009}.  
Hence, the interstellar grain size distribution will be very different in the dense phases of the interstellar medium than in the diffuse ISM and studies of the effects of grains on the opacity, ionization state and chemical inventory of molecular clouds will have to take this into account.

As \fg{tscales} illustrates, in our calculations, the average surface area of the grain mixture -- a proxy for the visual and near-IR extinction -- decreases by orders of magnitude during the initial coagulation process.  In a general sense this finding is in agreement with observational evidence for the importance of grain growth in molecular clouds as obtained from studies of dust extinction per unit column density of gas, where the latter is measured either through HI/H$_2$ UV absorption lines, sub-millimeter CO emission lines, or X-ray absorption \citep[cf.][]{Whittet:2005,Jura:1980,WinstonEtal:2007,GoldsmithEtal:1997}.  Obviously, this process is faster and therefore can proceed further in dense environments (\fg{tscales}).  As a corollary to this, the decrease in total surface area only occurs for timescales well in excess of the free-fall timescale. Hence, very short lived density fluctuations driven by turbulences will not show this effects of coagulation on the total grain surface area of dust extinction.

 The study by \citet{ChiarEtal:2007} is -- at first sight -- somewhat at odds with this interpretation. They find that the total near-IR extinction keeps rising when probing deeper into dense cores while the strength of the 10 $\mu$m feature abruptly levels off at a near-IR extinction value which depends somewhat on the cloud surveyed. The recent study by \citet{McClure:2009} also concludes that the strength of the 10 \micr\ feature relative to the local continuum extinction decreases dramatically when the K-band extinction exceeds 1 magnitude. Clearly, the two features are carried by different grain populations \citep{ChiarEtal:2007}. Indeed, models for interstellar extinction attribute the near-IR extinction to carbonaceous dust grains while the 10 $\mu$m feature is a measure of the silicate population \citep{DraineLee:1984}. Hence, these data suggest that silicates coagulate very rapidly when a certain density (\ie\ depth into the cloud) is reached -- essentially hiding silicates grains in the densest parts of the cloud from view -- while the carbonaceous grain population is not (as much) affected. In his study, \citet{McClure:2009} concludes that icy grains are involved in this change in extinction behavior with $A_K$. Likely, rather than the presence of the 13 \micr\ ice libration band affecting the silicate profile, this behavior reflects grain growth. Our study shows that coagulation in molecular clouds is greatly assisted by the presence of ice mantles. Once grains are covered by ice mantles, the increased `stickiness' of ice takes over and the precise characteristics of the core become immaterial. Perhaps, therefore, the data suggest that silicates rapidly acquire ice mantles while carbonaceous grains do not. However, there is no obvious physical basis for this suggestion. Further experimental studies on ice formation on different materials will have to settle this issue.

In this study we discussed observational implications of our model in a very coarse way, \ie\ by considering the reduction of the total geometrical surface area ($\kappa$) due to aggregation. We then find that its behavior can be largely expressed as function of the initial collision timescale, $t_\mathrm{coll,0}$.  However, for a direct comparison with observations, \eg\ the 10 \micr\ silicate absorption feature, it is relevant to calculate the extinction properties of the dust distribution as function of wavelength, and to assess, for example, the significance of porous aggregates \citep{MinEtal:2008,ShenEtal:2008}. This is the subject of a follow up study.

\section{Summary and conclusions}
\label{sec:concl}
We have studied the collisional growth and fragmentation process of dust in the environments of the molecular cloud (cores). In particular, we have focused on the collision model and the analysis of the several collisional evolution stages. Except for bouncing, the collision model features all relevant collisional outcomes (sticking, breakage, erosion, shattering). Furthermore, we have included off-center collisions in the recipe format and also prescribe the change to the internal structure in terms of the filling factor.  We have treated a general approach, and outcomes of future experiments -- either numerical or laboratory -- can be easily included.  The collision model features scaling of the results to the relevant masses and critical energies, which allows the calculation to proceeds beyond the sizes covered by the original numerical collision experiments. Our method is, in principle, also applicable to the dust coagulation and fragmentation stages in protoplanetary disks.

We list below the key results that follow from this study:
\begin{enumerate}
  \item The collisional evolution can be divided into three phases: \sumi\ $t<t_\mathrm{coll,0}$ in which the imprints of growth are relatively minor; \sumii\ a porosity-assisted growth stage, where the $N^2f(N)$ mass spectrum peaks at a well-defined size; and \sumiii\ a fragmentation stage, where the $N^2f(N)$ mass spectrum is relatively flat due to the replenishment of small particles by fragmentation.  Fragmentation is primarily caused by erosive collisions.
  \item A large porosity speeds up the coagulation of aggregates in the early phases. This effect is self-enhancing, because very porous particles couple very well to the gas, preventing energetic collisions capable of compaction. Growth in the second regime can therefore become very fast. Grazing collisions are largely responsible for obtaining fluffy aggregates in the first phases, further increasing the porosity.
  \item Silicate dust grains or, in general, grains without ice-coating are always in the fragmentation regime. This is a result of their relatively low breaking energy.  Freeze-out of ices, on the other hand, will significantly shift the fragmentation threshold upwards, fulfilling a prerequisite for significant aggregation in molecular clouds. 
  \item Likewise, the (monodisperse) grain size that enters the collision model is also critical for the strength of the resulting dust aggregates. Smaller grains will increase the strength significantly, due to increased surface contacts. Besides, a coagulation process that starts with small grains also results in the creation of very porous aggregates, which further enhances their growth. Although a single grain size cannot fully substitute for both the mechanical and aerodynamic properties of a grain size distribution, we have argued that for the MRN distribution a size of $0.1\ \micr$ reflects these properties best.
  \item If cloud lifetimes are restricted to free-fall times, little coagulation can be expected since the coagulation timescale is generally longer than $t_\mathrm{ff}$. However, if additional support mechanism are present, like ambipolar diffusion, and freeze-out of ice has commenced, dust aggregates of $\sim$100 \micr\ are produced, which will significantly alter the UV-opacity of the cloud. Conversely, our results reveal that the total dust surface area (and hence the extinction per H-nuclei) provides a convenient clock that measures the lifetime of a dense core in terms of the initial coagulation timescale. As observations typically reveal that the dust extinction per H-nuclei in dense cores has decreased substantially over that in the diffuse ISM, this implies that such cores are long-lived phenomena rather than transient density fluctuations.
  \item Despite the complexity of the collision model, we find that the decrease in (total) dust opacity can be expressed in terms of the initial collision time $t_\mathrm{coll,0}$ only, providing a relation for the density and lifetime of the cloud to its observational state (\fg{tscales}).
  \end{enumerate}

  \acknowledgement{The authors thank V.\ Ossenkopf for discussion on the results of his 1993 paper. C.W.O.\ appreciates useful discussions with Marco Spaans, which helped to clarify certain points of this manuscript. The authors also acknowledge the significantly contributions the referee, Vincent Guillet, has made to the paper by suggesting, for example, \se{sizea0}, \fg{fractal}, and \fg{sam}. These, together with many other valuable comments, have resulted in a significant improvement of both the structure and contents of the manuscript.}
\bibliographystyle{aa}
\bibliography{bibl}
\normalsize
\begin{appendix}
\section{Analytical background}
\label{app:app1}
\subsection{\label{app:relvel}Relative velocities and collision timescales of dust particles}
The friction time, $\tau_\mathrm{f}$, sets the amount of coupling between particles and gas. In molecular clouds the Epstein regime is applicable for which
\begin{equation}
  \tau_\mathrm{f} = \frac{3}{4\pi c_\mathrm{g} \rho_\mathrm{g}} \frac{m}{\sigma},
  \label{eq:tauf}
\end{equation}
where $m$ is the mass of the particle and $\sigma$ the average projected surface area.  For compact spheres \eq{tauf} scales linearly with radius, but for porous aggregates $\sigma$ can have a much steeper dependence on mass (in the case of flat structures, $\sigma \propto m$) and $\tau_\mathrm{f}$ a much weaker dependence.  For spherical grains of size $a_0$ and density $\rho_\mathrm{s}$ \eq{tauf} becomes
\begin{align}
  \label{eq:tau0}
  \tau_0 = & \tau_\mathrm{f}(a_0) = \frac{\rho_\mathrm{s} a_0}{c_\mathrm{g} \rho_\mathrm{g}} \\ 
         = & \nonumber  1.1\times10^2\ \mathrm{yr}\ \left( \frac{n}{10^5\ \mathrm{cm^{-3}}} \right)^{-1} \left( \frac{T}{10\ \mathrm{K}} \right)^{-1/2} \left( \frac{a_0}{0.1\ \mu\mathrm{m}} \right),
\end{align}
where $\rho_\mathrm{s}=2.65\ \mathrm{g\ cm^{-3}}$ is used, applicable for silicates. 

Any coagulation models requires the \textit{relative} velocities $\Delta v$ between two arbitrary particles. In turbulence, the motions of particles can become very correlated, though; \eg\ particles react in similar ways to the eddy in which they are entrained. The mean relative motion with respect to the gas, therefore, does not translate to $\Delta v$. \citet{VoelkEtal:1980}  have pioneered a study to statistically account for the collective effects of all eddies by dividing the eddies into two classes -- `strong' and `weak' -- depending on the turn-over time of the eddy with respect to the particle friction time. \citet{OrmelCuzzi:2007} approximated the framework of \citet{VoelkEtal:1980} and \citet{MarkiewiczEtal:1991} to provide closed-form expressions for the relative motion between two solid particles. In general three regimes can be distinguished:
\begin{enumerate}
  \item The low velocity regime, $\tau_2 \le \tau_1 \ll t_\mathrm{s}$. (Here, $\tau_1\ge\tau_2$ are the friction times of the particles). Relative velocities scale with the absolute difference in friction time, $\Delta v \propto \tau_1-\tau_2$.
  \item The intermediate velocity regime, for which $t_\mathrm{s} \ll \tau_1 \ll t_\mathrm{L}$. Particle velocities scale with the square root of the largest particle friction time. The particle motion will not align with eddies of shorter turn-over time. These `class II' eddies provide random kicks to the particle motion -- an important source for sustaining relative velocities of at least $\Delta v \sim v_\mathrm{s}$.
  \item The heavy particle regime, $\tau_1 \gg t_\mathrm{L}$, in which it is $\tau_2$ that determines the relative velocity.
\end{enumerate}

Comparing the friction time of the monomer grains (\eq{tau0}) with the smallest eddy turnover time, $t_\mathrm{s}$ (\eq{ts}), it follows that $\tau_0>t_\mathrm{s}$ under most molecular cloud conditions. We therefore focus on the intermediate velocity regime. In particular, the relative velocity between two equal solid spheres of $1<\tau_0/t_\mathrm{s}<\mathrm{Re}^{1/2}$ is \citep{OrmelCuzzi:2007} 
\begin{flalign}
  \label{eq:vturb0}
  \Delta v_0 \approx & \sqrt{3}v_\mathrm{s} \left( \frac{\tau_\mathrm{f}}{t_\mathrm{s}} \right)^{1/2} &\\ 
                   = & \nonumber 8.3\times10^2\ \mathrm{cm\ s^{-1}}\ \left( \frac{a_0}{0.1\ \mu\mathrm{m}} \right)^{1/2} \left( \frac{n}{10^5\ \mathrm{cm^{-3}}} \right)^{-1/4} \left( \frac{T}{10\ \mathrm{K}} \right)^{1/4}.&
\end{flalign}
Thus, velocities between silicate dust particles are $\sim$$10\ \mathrm{m/s}$, and have a very shallow dependence on density. The same expression holds when the silicates are coated with ice mantles that are not too thick, \ie\ $\rho_\mathrm{s}$ is still the silicate bulk density. Dust monomers then collide on a collision timescale of
\begin{flalign}
  \label{eq:tcoll}
  t_\mathrm{coll,0} =& \left( n_\mathrm{d} \Delta v_0 4\pi a_0^2 \right)^{-1} = \frac{\rho_\mathrm{s} a_0 {\cal R}_\mathrm{gd}}{3\rho_\mathrm{g} \Delta v_0} & \\ 
                    =& \nonumber 8.5\times10^4\ \mathrm{yr}\ \left( \frac{a_0}{0.1\ \mu\mathrm{m}} \right)^{1/2} \left( \frac{n}{10^5\ \mathrm{cm^{-3}}} \right)^{-3/4} \left( \frac{T}{10\ \mathrm{K}} \right)^{-1/4},&
\end{flalign}
where $n_\mathrm{d}$ is the dust number density and ${\cal R}_\mathrm{gd}=100$ is the standard gas-to-dust density ratio by mass.  

\Eqs{vturb0}{tcoll} are only valid for solid particles. However, we can scale these relations to two arbitrary but equal aggregates of filling factor $\phi_\sigma$ and (dimensionless) mass $N$. Because $m\propto N$ and $\sigma \propto (N/\phi_\sigma)^{2/3}$ it follows that
\begin{subequations}
  \label{eq:simple}
  \begin{equation}
    \tau_\mathrm{f} = N^{1/3} \phi_\sigma^{2/3} \tau_0;
  \end{equation}
  \begin{equation}
    \Delta v \simeq \left( \frac{\tau_\mathrm{f}}{\tau_0}\right)^{1/2} \Delta v_0 = N^{1/6} \phi_\sigma^{1/3} \Delta v_0;
  \end{equation}
  \begin{equation}
    t_\mathrm{coll} = \left( n_\mathrm{d} \Delta v \sigma^\mathrm{C} \right)^{-1} \simeq N^{1/6} \phi_\sigma^{1/3} t_\mathrm{coll,0},
    \label{eq:tcoll-simple}
  \end{equation}
\end{subequations}
where in the latter equation we substituted for simplicity the geometrical cross section $\sigma$ for the collisional cross-section $\sigma^\mathrm{C}$ and used the monodisperse assumption $n_\mathrm{d} \propto N^{-1}$ for the dust number density $n_\mathrm{d}$ (this is of course only applicable for narrow distributions). Thus, \eq{tcoll-simple} shows that the collision timescale decreases for very porous particles with low filling factors.

\subsection{\label{app:simple-anal}A simple analytical model for the initial stages of the growth}
Despite the complexity of the recipe, it is instructive to approximate the initial collisional evolution of the dust size distribution with a simple analytical model \citep[\cf][]{Blum:2004}.  \Fg{por-analys} suggests that the initial evolution of $\phi_\sigma$ can be divided in two regimes, where the transition point occurs at a mass $N_1$.  Initially ($N<N_1$), the filling factor is in the fractal regime, which can be well approximated by a power-law, $\phi_\sigma\simeq N^{-3/10}$. We refer to the fractal regime as including hit-and-stick collisions (no restructuring) as well as collisions for which $E>5E_\mathrm{roll}$ but which do not lead to \textit{visible} restructuring, \ie\ only a small fraction of the grains take part in the restructuring.  For $N>N_1$ the filling factor starts to flatten-out. It is difficult to assign a trend for $\phi_\sigma$ in the subsequent evolution. Following \fg{por-analys} we may assume that initially $\phi_\sigma$ stays approximately constant for several orders of magnitude in $N$, although at some point it will quickly assume its compact value of 33\%.  A sketch of the adopted porosity structure and the resulting scaling of velocities and timescales is presented in \fg{sketch}, in which it is assumed that the collapse of the porous structure takes place \textit{after} the point where the first erosive collisions occurs, at $N=N_2$. 

\begin{figure}
  \centering
  \includegraphics[width=80mm]{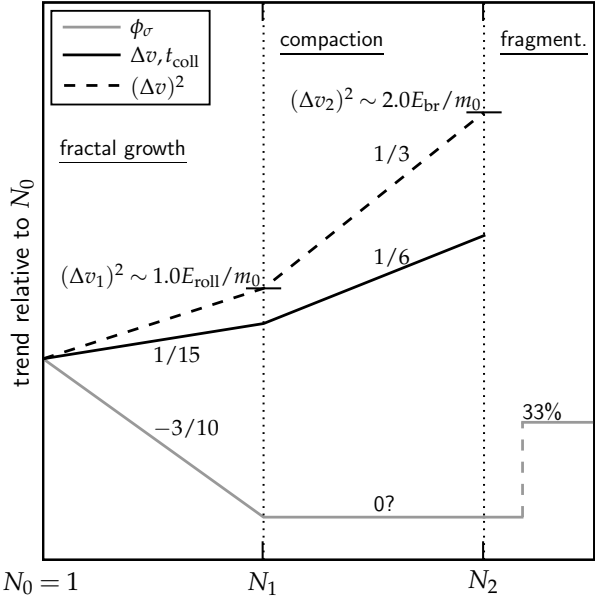}
  \caption{\label{fig:sketch} (\textit{gray solid line}) A simplified model for the behavior of the filling factor with growth. Initially, for $\Delta v_0 < \Delta v_1$, the porosity decreases (fractal growth regime). This phase is followed by a `status quo' phase where filling factors will be approximately constant. The first compaction event is reached when velocities reach $\Delta v_1$ and fragmentation sets in when relative velocities exceeds $\Delta v_2$. (\textit{black solid line}) Trend of the collision velocity and collision timescale. (\textit{dashed line}) Trend of $(\Delta v)^2$. The numbers denote the power-law exponents.}
\end{figure}
From the collision recipe (\se{colrec}) we identify the critical energies at which visible compaction and fragmentation occur. Compaction requires collisions between similar size particles (\ie\ the global recipe) and \fg{global-recipe} shows that the transition ($C_\phi>1$) occurs at an energy of $E/N_\mathrm{tot} E_\mathrm{roll} \simeq 0.2$. Regarding fragmentation, the simulations clearly show that small particles are replenished in the form of erosive collisions (local recipe). From \fg{local-recipe} we assign an energy threshold of $E/N_\mu E_\mathrm{br} \simeq 1.0$. Working out these expressions and using a typical mass ratio of 3 for the global recipe ($N_\mu = N/6$), the corresponding critical velocities become $(\Delta v_1)^2 \simeq 1.0E_\mathrm{roll}/m_0$ and $(\Delta v_2)^2 \simeq 2.0E_\mathrm{br}/m_0$, respectively. These energy thresholds are also indicated in \fg{sketch}.

From these expressions and the initial expressions for the relative velocity and the collision timescale (\eqs{vturb0}{tcoll}), the turn-over points $N_1$ and $N_2$ can be calculated. We assume that $\Delta v_0 < \Delta v_1$ such that a fractal growth regime exist. Then, the first transition point is reached at a mass
\begin{flalign}
\label{eq:N1}
  N_1 & \sim \left( \frac{\Delta v_1}{\Delta v_0} \right)^{15} = \left( \frac{1.0E_\mathrm{roll}}{m_0 (\Delta v_0)^2} \right)^{7.5} = &\\
      & = 2\times10^3 \left( \frac{n}{10^5\ \mathrm{cm^{-3}}} \right)^{3.75} \left( \frac{\gamma}{370\ \mathrm{erg\ cm^{-2}}} \right)^{7.5} \left( \frac{a_0}{0.1\ \micr} \right)^{-22.5}. \nonumber &
\end{flalign}
Unfortunately, the high powers make the numeric evaluation rather unstable.  In our simulations we find that $N_1 \sim 10^4$. Subsequently, we can write for the second transition point, the onset of fragmentation, $N_2$,
\begin{flalign}
  \label{eq:N2}
  \frac{N_2}{N_1} & \sim \left( \frac{\Delta v_2}{\Delta v_1} \right)^6 = \left( 2.0 \frac{E_\mathrm{br}}{E_\mathrm{roll}} \right)^3 & \\
                  &  = 5\times10^4 \left( \frac{\gamma}{370\ \mathrm{erg\ cm^{-2}}} \right)^2 \left( \frac{a_0}{0.1\ \micr} \right) \left( \frac{ {\cal E}^\star}{3.7\times10^{10}\ \mathrm{dyn\ cm^{-2}}} \right)^{-2},\nonumber &
\end{flalign}
which corresponds also well to the results from the simulation for which $N_2 \sim 10^8$. In our simulations the \textit{first} fragmentation involves particles that are still relatively porous, such that the assumption in \fg{sketch} about the porosity of the $N_2$-particles is justified. However, once steady-state has been reached, particles of $N_2\sim10^8$ will have a 33\% filling factor (see \fg{por-analys}).

\begin{figure}
  \centering
  \includegraphics[width=80mm]{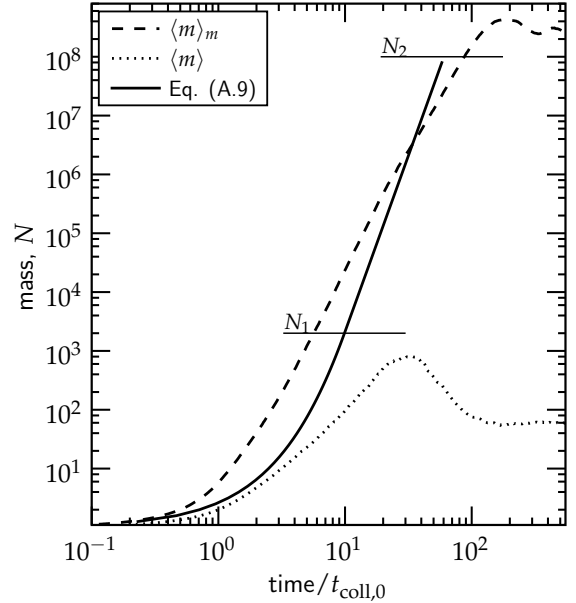}
  \caption{\label{fig:sam}Results of the simple analytic model (\textit{solid line}) and comparison to the $\langle m \rangle$ and $\langle m \rangle_m$ statistics of the numerical results of our standard model. }
\end{figure}
Using again the monodisperse assumption we also obtain the timescales $t_1,t_2$ at which these transition points are reached. Writing, 
\begin{equation}
  \frac{\mathrm{d}N}{\mathrm{d}t} = \frac{N}{t_\mathrm{coll}} = \frac{N^{5/6} \phi_\sigma^{-1/3}}{t_\mathrm{coll,0}},
\end{equation}
where \eq{tcoll} is inserted for $t_\mathrm{coll}$, we insert the power-law dependence of $\phi_\sigma$ on $N$ to obtain $t$.  Straightforward integration gives
\begin{equation}
  \frac{t}{t_{\mathrm{coll},0}} = \int_1^{N_1} N'^{-14/15} \mathrm{d}N' + N_1^{-1/10} \int_{N_1}^{N_2} N'^{-5/6} \mathrm{d}N',
  \label{eq:t2}
\end{equation}
see \fg{sam}, where we plotted the results of \eq{t2} together with the averaged mass and the mass-averaged mass of the distribution of the standard model.  It follows that the fractal growth stages takes $\sim$10 $t_\mathrm{coll,0}$, or $\sim$$8\times10^5\ \mathrm{yr}$ (\cf\ $\sim$$6\times10^5\ \mathrm{yr}$ in the simulation), while $N_2$ is reached at $\sim$$60\ t_\mathrm{coll,0}$ (\cf\ $\sim$$30 t_\mathrm{coll,0}$ in the simulation). At larger time our fit may overestimate the growth rates somewhat because it assumes the filling factor stays fixed at its low value. Overall, the model nicely catches the trends of the growth and is also in good agreement with previous approaches \citep{Blum:2004}, although, being a monodisperse model, it cannot fit both $\langle m \rangle$ and $\langle m \rangle_m$.

\section[The numerical collision experiments]{The numerical collision experiments}
\label{app:app2}
The skeleton of our collision model consists of the outcomes of many aggregate-aggregate collision simulations. In this appendix we briefly review the setup of the simulations (\app{collision-setup}), discuss some auxiliary relations required to complete the collision model (\app{auxil}), discuss the output format of the binary collision model (\app{colltab}), and present a few limitations that arise due to our reliance on the outcomes of the numerical simulations (\app{merits}).

\subsection{\label{app:collision-setup}Collision setup and input parameters}
Collisions between aggregates are modeled using the soft aggregates numerical dynamics (SAND) code \citep{2002Icar..157..173D,PaszunDominik:2008}. This code treats interactions between individual grains held together by surface forces in a contact area \citep{1971ProcRSocLonA..324.301J,1975JournCollInterfSci...53..314D}. The SAND code calculates the equation of motion for each grain individually and simulates vibration, rolling, twisting, and sliding of the grains that are in contact.  These interactions lead to energy dissipation via different channels. When two grains in contact are pulled away, the connection may break, causing loss of the energy. Monomers may also roll or slide over each other, through which energy is also dissipated \citep{1995PhMA...72..783D,1996PhMA...73.1279D,DominikTielens:1997}.  For further details regarding this model and testing it against laboratory experiments we refer the reader to \citet{PaszunDominik:2008}.

\begin{figure}
  \includegraphics[width=88mm]{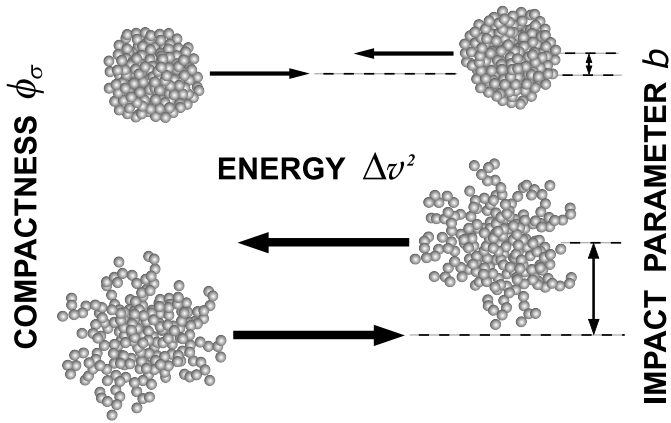}
  \caption{Sketch of the initial setup of our simulations. The key input parameters $\Delta v$, $b$, and $\phi_\sigma$ are illustrated.}
  \label{fig:sketch-parameter-space}
\end{figure}
To provide both a qualitative and a quantitative description of a collision, \citet{PaszunDominik:2009} have performed a large number of simulations, exploring an extensive parameter space. They formulate a simple collision recipe that quantifies how kinetic energy, compactness, and mass ratio affect the outcome of aggregate-aggregate collisions. These results are presented in tabular format (\app{colltab}). \Fg{sketch-parameter-space} sketches the setup of these numerical experiments, illustrating three parameters that shape the outcome of a collision: the initial compactness as represented by the \emph{geometrical} filling factor $\phi_\sigma$ (see below, \eq{phi-geom-def}), the collision energy $E$, and the impact parameter $b$.  A collision for each parameter set is repeated six times at different orientations, providing information on the range of outcomes. Because of the occasionally fluffy structure of the aggregates not all orientations result in a collision hit, especially not those at large impact parameter. An overview of the parameter ranges is given in \Tb{params-ranges}. The radius of the monomer grains in the simulation is $a_0 = 0.6\ \mu\mathrm{m}$ and the adopted material properties reflect silicates ($\gamma = 25\ \mathrm{erg\ cm^{-2}}$, ${\cal E} = 2.8\times10^{11}\ \mathrm{dyn\ cm^{-2}}$). 

\begin{table}
  \begin{center}
  \caption{\label{tab:params-ranges}Parameters used in the numerical simulations. }
  \begin{tabular}{rrrr}
    \hline\hline
     $\Delta v$ & $b/b_\mathrm{max}$ & $\phi_\sigma$  & $N_2/N_1$ \\ 
   \strut [m/s] &                    &                & \\
     (1)        & (2)                & (3)            & (4) \\
    \hline
    0.05 & 0.0   & 0.070 & 1.0   \\
    0.30 & 0.25  & 0.090 & $10^{-3}$   \\
    0.50 & 0.5   & 0.122 &  \\
    0.75 & 0.75  & 0.127 &  \\
     1.0 & 0.875 & 0.155 &   \\
     2.0 & 0.95  & 0.161 &   \\
     4.0 &       & 0.189 &   \\
     6.0 &       & 0.251 &   \\
     8.0 &       &       &   \\
    10.0 &       &       &   \\ \hline
  \end{tabular}
  \end{center}
  {\small \textsc{Note.} (1) relative velocity; (2) impact parameter normalized to the sum of the outer radii $b_\mathrm{max}$; (3) geometrical filling factor; (4) mass ratio.}
\end{table}
Relative velocities $\Delta v$ are chosen such that all relevant collision regimes are sampled:  from the pure hit-and-stick collisions, where particles grow without changing the internal structure of the colliding aggregates, up to catastrophic destruction, where the aggregate is shattered into small fragments. In the intermediate energy regime, restructuring without fragmentation takes place. For aggregate collisions at large size ratios, high velocity impacts result in erosion of the large aggregates.

The impact parameter $b$ is also well sampled. We probe central collision ($b=0$), where aggregates can be compressed, grazing impacts ($b\approx b_\mathrm{max}$), where particles can be stretched due to inertia, and several intermediate cases. In \Tb{params-ranges} the impact parameter is defined relative to the outer radius of the particles, $b_\mathrm{max}=a_{\mathrm{out},1}+a_{\mathrm{out},2}$. Here the outer radius $a_{\mathrm{out}}$ is the radius of the smallest sphere centered at the center-of-mass of the particle that fully encloses it. In the \citet{PaszunDominik:2009} study the outcomes of a collision are averaged over the impact parameter $b$. However, in a Monte Carlo treatment, the averaging over impact is not necessary, and the normalized impact parameter $\tilde{b}=b/b_\mathrm{max}$ can be determined using a random deviate $\overline{r}$, \ie\ $\tilde{b}=\overline{r}^{1/2}$. We have chosen to use the raw data sampled by the \citet{PaszunDominik:2009} parameter study, explicitly including $\tilde{b}$ as an input parameter for the Monte Carlo model.  In this way the effects of off-center impacts can be assessed, \ie\ by comparing it to models that contain only head-on collisions.

The internal structure of the aggregates, or initial compactness, is characterized by the geometrical filling factor, $\phi_\sigma$ (\eq{phi-geom-def}).  To obtain $\phi_\sigma$, the projected surface area, $\sigma$, is averaged over a large number of different orientations of the particle. The parameter space of the filling factor $\phi_\sigma$ is chosen such that we sample very porous, fractal aggregates that grow due to the Brownian motion \citep{BlumSchraepler:2004,PaszunDominik:2006}, through intermediate compactness aggregates that form by particle-cluster aggregation (PCA), up to compact aggregates that may result from collisional compaction. 

The final parameter that determines a collision outcome is the mass ratio, $N_2/N_1$ ($N_1$ being the larger aggregate).  The \citet{PaszunDominik:2009} experiments sample a mass ratio between 1 and $10^{-3}$. In this study, however, we will only use the collisions corresponding to the two extreme values (\ie\ mass ratios of 1 and $10^{-3}$) as representatives of two distinct classes: \textit{global} and \textit{local}.  

\subsection{Auxiliary relations for the collision recipe}
There are a few more relations required for a consistent approach to the collision recipe. These are presented here for completeness.

\label{app:auxil}
\subsubsection{\label{sec:ff-small}The filling factor of small fragments}
A common filling factor can be assigned to the small fragments produced by erosive or fragmenting collisions that constitute the power-law component. The compactness of these particles depends only on mass and is presented in \fg{small-phisigma}, where fragments produced in many simulations, reflecting a variety of collision properties, are plotted. Almost all particles are located along the power-law with the slope of $-0.33$. This provides an easy prescription for the filling factor of small fragments. The dependence indicates a fractal structure (with the fractal dimension of about $D_\mathrm{f} \approx 2.0$) of aggregates formed in a fragmentation event, since non-fractal aggregates would have a filling factor independent of mass.

\begin{figure}
  \includegraphics[width=\figwidth,clip]{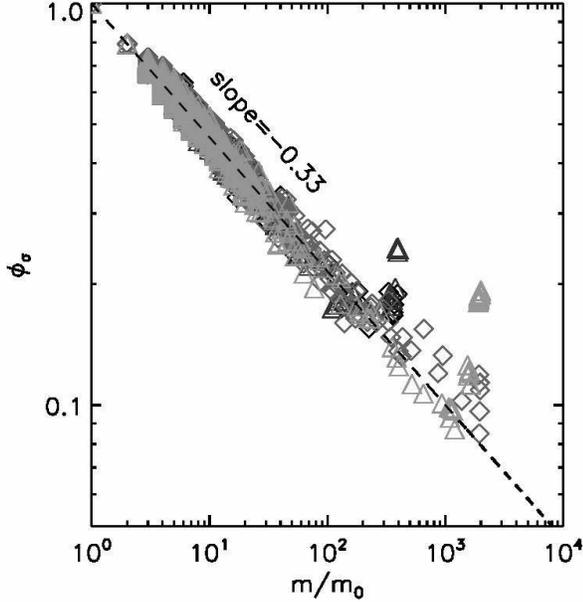}
  \caption{\label{fig:small-phisigma}The geometrical filling factor as a function of fragment mass. Many simulations with different sets of parameters are overplotted. The dashed line indicates the least square power-law fit, $\phi_\sigma \simeq (m/m_0)^{-0.33}$.}
\end{figure}
As shown by \citet{PaszunDominik:2009}, after reaching the maximum compaction, further increase of the impact energy produces more restructuring and results in a flattening of the produced aggregate. Therefore, very fluffy particles can be produced in collisions of massive aggregates, where the power-law component extends to larger $N$.  This behavior is also observed in \fg{small-phisigma}, where fluffy, small fragments follow the power-law relation, while some large, still compact particles are above the dashed line.

\subsubsection{\label{app:aout-asig}Relation between $a_\mathrm{out}$ and $a_\sigma$}
In this study we characterize aggregates by two different radii: the outer radius $a_\mathrm{out}$ and the projected surface equivalent radius $a_\sigma$.  The first is used as a reference to the impact parameter $b$, \ie\ the collision offset is determined relative to the largest impact parameter, $b_\mathrm{max} = a_{\mathrm{out},1} + a_{\mathrm{out},2}$. The cross-section equivalent radius $a_\sigma$ defines our structural parameter $\phi_\sigma$ (see \eq{phi-geom-def}). We determine the relation between the two radii ($a_\mathrm{out}$ and $a_\sigma$) empirically. Both $a_\mathrm{out}$ and $a_\sigma$ are determined for many aggregates of various shape and mass. We sample particles with the fractal dimension in the range of $D_\mathrm{f}=1\ \ldots\ 3$ and masses from several to a few thousands monomer masses. These aggregates were produced using an algorithm developed by \citet{2000JournCollInterfSci...229..261F}.

\begin{figure}
  \includegraphics[width=\figwidth,clip]{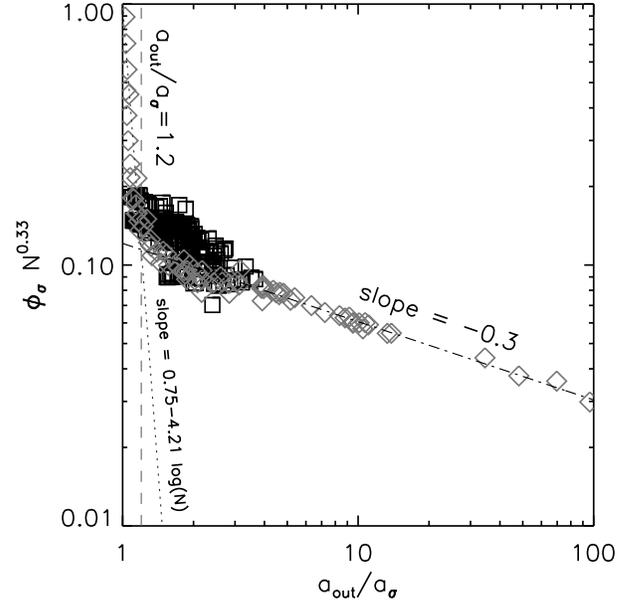}
  \caption{ \label{fig:aout-asig}The geometrical filling factor dependence on the ratio of outer to geometrical radii. In this figure we plot $\phi_\sigma N^{0.33}$ to scale the data for aggregates of different mass.} 
\end{figure}
\Fg{aout-asig} shows the filling factor determined for different aggregates versus the ratio of the outer radius over the cross-section equivalent radius.   Diamonds correspond to the produced aggregates of different fractal dimension and mass. The mass dependence in \fg{aout-asig} is taken into account by plotting $\phi_\sigma \ N^{0.33}$. In this way the data for all aggregates are well confined along a single curve. At small $a_\mathrm{out}/a_\sigma$ the curve decreases very steeply with increasing $a_\mathrm{out}/a_\sigma$. This corresponds to compact particles for which $a_\mathrm{out}/a_\sigma$ is insensitive to filling factor. The line, however, breaks at about $a_\mathrm{out}/a_\sigma \approx 1.2$ and turns in to a power-law with a slope of $-0.3$. This shallow relation represents fluffy aggregates that show a large discrepancy between the projected surface equivalent radius and the outer radius.

In order to provide a simple relation between the two radii, two power-law functions are fitted to the two regimes: compact particles below $a_\mathrm{out}/a_\sigma=1.2$ and fluffy aggregates above that limit.  These two functions are given by
\begin{subequations}
  \label{eq:psig-arat-fits}
  \begin{equation}
    \phi_\sigma^\mathrm{compact} =
    \left(\frac{a_\mathrm{out}}{a_\sigma}\right)^{0.75 - 4.21 \log N}
    \label{eq:psig-arat-fit-c}
  \end{equation}
  \begin{equation}
    \phi_\sigma^\mathrm{fluffy} = 1.21
    \left(\frac{a_\mathrm{out}}{a_\sigma}\right)^{-0.3} N^{-0.33}.
    \label{eq:psig-arat-fit-f}
  \end{equation}
\end{subequations}
To further verify these relations we use particles produced in several simulations performed by \citet{PaszunDominik:2009}. These aggregates are indicated in \fg{aout-asig} by black squares. They show a similar relation to the one obtained in \eq{psig-arat-fits}. Points that are slightly shifted above the fitted lines correspond to aggregates that are partly compressed (they did not reach the maximum compaction). Their compact cores are still surrounded by a fluffy exterior that causes a small increase of the ratio of the outer radius over the projected surface equivalent radius $a_\mathrm{out}/a_\sigma$. This behavior, however, occurs at a relatively small value of $a_\mathrm{out}/a_\sigma < 2$. At a larger size ratio the filling factor falls back onto the power-law given in \eq{psig-arat-fit-f}.

We remark here that, although the fits present a general picture, situations where $a_\mathrm{out} \gg a_\sigma$ are not likely to materialize when $N\gg1$. This would corresponds to very open fractals of fractal dimension less than two. Instead, in our models $\phi_\sigma N^{0.33} \gtrsim 0.1$ and we therefore always have $a_\mathrm{out} \sim a_\sigma$. Consequently, the fraction of missing collisions $f_\mathrm{miss}$ is close to zero in most of the cases. 
\subsubsection{\label{app:hitandstick}Hit and stick}
At very low energies ($E \le 5E_\mathrm{roll}$) two aggregates will stick where they meet, without affecting the internal structure of the particles. This is the `hit-and-stick' regime in which the collisional growth can often be described by fractal laws. Two important limits are cluster-cluster coagulation (CCA) and particle-cluster coagulation (PCA). In the former, two particles of equal size meet, which often results in very fluffy structures, whereas PCA describes the process in which the projectile particles are small with respect to the target. The filling factor then saturates to a constant value. In the case of monomers, the filling factor will reach $15\%$ \citep{1992A&A...263..423K}.

In general particles do not merely collide with either similar-size particles or monomers.  Every size-ratio is possible and leads to a different change in filling factor. \citet{OrmelEtal:2007} provide an analytical expression, based upon fits to collision experiments of \citet{Ossenkopf:1993}, that give the increase in void space as function of the geometrical volume of the collision partners. Here, the geometrical volume $V$ is the volume that corresponds to the geometrical radius, $a_\sigma$.  In this study additional numerical collision experiments were used to further constrain these analytical fits. These experiments involved several `monomer-bombardments' of aggregates of different initial filling factor. Using these data, we fit the volume increase as  
\begin{align}
  \nonumber
  \frac{V_\mathrm{void}}{V_0} = \textrm{max} & \left[ (V_1+V_2) \left( \left(1+\frac{V_2}{V_1}\right)^{3\delta/2-1} -1 \right), \right. \\ 
                                             & \left. \frac{N_2}{0.087\phi_2} \exp\left[-\left(\frac{15V_2}{V_1}\right)^{0.25}\right] \right] ,
  \label{eq:hs}
\end{align}
where $V_\mathrm{void}$ is the increase in void space (leading to a lower $\phi_\sigma$), $V_1>V_2$ the geometrical volumes of the collision partner, $N_2$ the number of grains in the smaller particle, and $V_0$ the monomer volume. The first term converges to CCA in the limit of $V_2=V_1$, and is the same as in \citet{OrmelEtal:2007}. Here, $\delta=0.95$ is an exponent that reflects the fractal growth in this limit \citep{Ossenkopf:1993}. The second expression converges to PCA in the limit of $V_2 \ll V_1$. The rationale of providing a second expression is that in the case of $V_2 \ll V_1$ (PCA) the first expression goes to zero very quickly (no voids are added), which is inconsistent with the PCA limit of 15\%. From the results of our new collision experiments we have introduced an exponent of $0.25$ to the PCA-part of \eq{hs}, which softens the decrease of $V_\mathrm{void}$ with increasing mass ratio.

However, not all numerical experiments could be fitted equally well. In fact, we had to compromise. It is likely that a better fit involves more parameters, \eg\ the elongation of the aggregates or their fractal dimension. Here, we have adopted approximate fits that follow the qualitative picture in both the CCA ($V_1 = V_2$) and the PCA ($V_2 \ll V_1$) limit.  Remark, finally, that for the molecular cloud environment the hit-and-stick regime is only relevant in the initial stages of coagulation at densities of $n\ge10^5\ \mathrm{cm^{-3}}$ or grain sizes $a_0 < 0.1\ \mu\mathrm{m}$.

\subsection{The collision tables}
\label{app:colltab}
Given the level of complexity, it is not feasible to provide simple analytical expressions for the collision outcome (in terms of the parameters listed in \Tb{recipe-quantities}) as function of the collision parameters ($E,\phi_\sigma,\tilde{b},N_1/N_2$). Therefore, like in \citet{PaszunDominik:2009}, the results are expressed in a tabular format. In total 72 tables are provided. They describe six output quantities (see \Tb{recipe-quantities}) for six impact parameters $b$ and for both the local and the global recipes. Since listing all these tables here is impractical, we will provide them in the digital form as \textit{online material} accompanying this manuscript. We present two examples to illustrate the format.

Each table lists one output quantity as function of the dimensionless energy parameter $\varepsilon$ and the initial filling factor of aggregates $\phi_\sigma$. The only exception concerns the fraction of missed collisions, $f_\mathrm{miss}$. This quantity provides a correction to the collision cross-section of particles, in our case calculated from the outer radius of an aggregate $a_\mathrm{out}$ (cf.\ \app{mccycle}). The filling factor $\phi_\sigma$ is not an appropriate quantity to use here, because it is ambiguous where it concerns the structure of particles. For example, low $\phi_\sigma$ could mean either a very fractal structure (and correspondingly high number of missing collisions) or a porous but homogeneous structure (and low number of missing collisions). Therefore, it is more appropriate to relate the probability of a collision miss to the radii with which the particle is characterized.  Thus, $f_\mathrm{miss}$ is provided as a function of the ratio of the outer radius over the projected surface equivalent radius, $a_\mathrm{out}/a_\sigma$. 

\begin{table}
  \centering
  \caption{\label{tab:rec-tab-fpwl}Example of an output table from the online data ($f_\mathrm{pwl}$ at $b=0$ in the global recipe).} 
  \begin{tabular}{lllll}
    \hline \hline
    \multicolumn{1}{c}{$\varepsilon$} & \multicolumn{4}{c}{$\phi_\sigma^\mathrm{ini}$} \\
    \cline{2-5}
                   &    0.1219    &   0.1553      &   0.1893    &    0.2505 \\
    \hline                         
    $5.721(-4)$    &     0.00000  &      0.000    &     0.000 &     0.000\\
    $2.595(-2)$    &     0.00000  &     2.500(-3) &     0.000 &     0.000\\
    $5.721(-2)$    & $8.330(-4)$  &     0.000     &     0.000 &     0.000\\
         0.1287    & $9.250(-2)$  &     3.042(-2) &     8.750(-3) & 2.917(-2)\\
         0.2288    &    0.3888    &     9.417(-2) &     3.042(-2) & 1.500(-2)\\
         0.9153    &    0.9575    &    0.6033     &    0.2438 &     0.1158\\
          3.661    &     1.000    &     1.000     &     1.000 &     0.7271\\
          8.238    &     1.000    &     1.000     &     1.000 &     1.000\\
          14.65    &     1.000    &     1.000     &     1.000 &     1.000\\
    \hline
  \end{tabular}
\end{table}
Each table is preceded by a header that specifies: the corresponding recipe (keyword: \textsc{global} or \textsc{local}), the corresponding impact parameter $b$, and the quantity listed in the table (keywords are: \textit{fmiss}, \textit{Nf}, \textit{Sf}, \textit{fpwl}, \textit{q}, \textit{Csig}). In the case of \Tb{rec-tab-fpwl} the header is
\begin{verbatim}
# GLOBAL, b=0.0, Q=fpwl
\end{verbatim}
Therefore, \Tb{rec-tab-fpwl} presents the fraction of mass in the power-law component, $f_\mathrm{pwl}$, for the global recipe and for head-on collision. 

In each table the first column and the first row specify the normalized energy parameter $\varepsilon$ and the initial filling factor $\phi_\sigma^\mathrm{ini}$ (or the ratio of the outer over the geometrical radii $a_\mathrm{out}/a_\sigma$ in the case of $f_\mathrm{miss}$), respectively. Here, $\varepsilon$ denotes the collision energy scaled by a normalization constant that involves \sumi\ the breaking \textit{or} rolling energy and \sumii\ the reduced \textit{or} total number of particles, see \se{par-space-normalization} and \Tb{recipe-quantities}. In the case of \Tb{rec-tab-fpwl} the scaling parameter is $\varepsilon=E/E_\mathrm{br}N_\mathrm{tot}$.

\Tb{rec-tab-fpwlloc} is the second example. It is taken from the local recipe and it presents the $f_\mathrm{pwl}$ quantity for the head-on collision. The dimensionless energy parameter $\varepsilon$ has fewer entries in the local recipe tables than in the global recipe. In \Tb{rec-tab-fpwlloc} the energy is scaled by reduced number of monomers $N_\mu$ (local recipe scaling) and by the breaking energy $E_\mathrm{br}$ (erosion scaling) as indicated in \Tb{recipe-quantities}. The header in this case is
\begin{verbatim}
# LOCAL, b=0.0, Q=fpwl
\end{verbatim}
 \begin{table}
   \centering
   \caption{\label{tab:rec-tab-fpwlloc}Example of an output table from the online data ($f_\mathrm{pwl}$ at $b=0$ in the local recipe).}
   \begin{tabular}{lllll}
     \hline \hline
     \multicolumn{1}{c}{$\varepsilon$} & \multicolumn{4}{c}{$\phi_\sigma^\mathrm{ini}$} \\
     \cline{2-5}
              &   7.009(-2) &  9.047(-2) &    0.1268 & 0.1610   \\
     \hline         
     0.2288   &   0.000   &     0.00000 &     0.0000 & 0.000 \\
     0.9154   &   1.001   &      0.3337 &     0.0000 & 0.000 \\
      3.661   &   4.004   &       46.55 &     6.340  & 0.667 \\
      14.65   &   7.007   &       67.07 &     35.20  & 9.009 \\
      32.95   &   7.508   &       148.3 &     58.22  & 16.02 \\
      58.58   &   9.510   &       129.2 &     62.40  & 30.03 \\
      \hline
\end{tabular} 
\end{table}
Note that in the local recipe the filling factors are lower. In this case larger aggregates are used to model collisions at large mass ratio, $N_1/N_2 = 10^3$. The fractal structure of these aggregates results in a lower filling factor. 

\subsection{Limitations of the collision recipe}
\label{app:merits}

The main limitation of the collision recipe is that, due to computational constraints, the binary aggregate simulations can only simulate aggregates of a mass $N\lesssim10^3$.  For the recipe to become applicable for large aggregates \textit{scaling} of the results of the collision experiments is required.  This is a critical point of the recipe for which suitable dimensionless quantities had to be determined.  
However, the extrapolation assumes that the collision physics that determines the outcomes of collisions at low-$N$ also holds at large scales. This is a crucial assumption in which collisional outcomes like bouncing are \textit{a priori} not possible because these do not take place at the low-$N$ part of the simulations. 

Bouncing of aggregates is observed in laboratory experiments \citep{BlumMuench:1993,2006AdPhy..55..881B,Langkowski:2008,WeidlingEtal:2009}, whereas it does not occur in our simulations.  For silicates, bouncing occurs at sizes above approximately $100\ \mu$m (\ie\ $N>10^9$ particles) and is not fully understood from a microphysical perspective.  In the case of ice-coated silicate grains, which provide stronger adhesion forces, our simulations show that growth proceeds to $\sim$$100\ \micr$ sizes.  In this case, therefore, bouncing might slow down the growth earlier than observed in our experiments, especially when the internal structure has already re-adjusted to a compact state. However, it is presently unclear how these laboratory experiments apply to ice aggregates and hence whether and to what extent the results would be affected by bouncing. We recognize that this may, potentially, present a limitation to the growth of aggregates in molecular clouds, but also emphasize it will not affect the main conclusions from this study as in only a few models aggregates grow to sizes $\gg$100 $\micr$.

Another assumption of the collision model is that the grains have a spherical geometry. Again, computational constraints rule out numerical modeling of randomly shaped particles.  Whether erratically shaped grains would help or harm the sticking or bouncing is unclear. Because the strength of an aggregate is determined by the amount of contact area between two grains, the strength of irregularly shaped monomers depends on the local radius of curvature. Therefore, highly irregular grains are held by contacts of much smaller size, because they are connected by surface asperities. On the other hand, irregular grains may form more than one contact. However, the geometry of the grains does not necessarily pose a bottleneck to the validity of the collision model. Instead, like the size distribution, the consequence of irregularly shaped monomers is reflected in a different energy scaling.

\section{The Monte Carlo program}
\label{app:mccycle}
The advantage of a Monte Carlo (MC) approach to the calculation of the collisional evolution is that collisions are modeled individually and that they, therefore, bear a direct correspondence to the collision model.  Furthermore, in a MC approach structural parameters (like $\phi_\sigma$) can be easily included and the collisional outcome can be quantified in detail.  From the two particle properties ($N$ and $\phi_\sigma$) the collisional quantities are derived, \eg\ the relative velocities $\Delta v$ between the aggregates (see \app{relvel}). Then, from $\Delta v$ and the particle's outer radii we calculate the collision rates between all particles present in the MC simulation. After the MC model has selected the collision partners, the collision recipe is implemented.  First, the particle properties ($m,\phi_\sigma$) and the collision properties ($\Delta v$) are turned into a collision `grid point' given by the dimensionless $\varepsilon, \phi_\sigma$ and $\tilde{b}$.  The six collision quantities (\Tb{recipe-quantities}) are then taken from the appropriate entries from the recipe tables.  Finally, these quantities specify the change to the initial particle properties ($m,\phi_\sigma$) and also describe the properties of the collision fragments.

By virtue of the scaling relations discussed in \se{par-space-normalization} the MC coagulation model is able to treat much larger aggregates than the binary collision experiments. A broad size distributions, which may, \eg\ result from injection of small particles due to fragmentation, can, however, become problematic for a MC approach, since the high dynamic range required consumes computational resources.  To overcome this problem we use the grouping method outlined by \citet{OrmelSpaans:2008}. In this method the 1-1 correspondence between a simulation particle and a physical particle is dropped; instead, the simulated particles are represented by \textit{groups} of identical physical particles. The group's mass is determined by the peak of the $m^2f(m)$ mass distribution -- denoted $m_\mathrm{p}$ -- and particles of smaller mass `travel' together in groups of total mass $m_\mathrm{p}$. Grouping entails that a large particle can collide with many small particles simultaneously -- a necessary approximation of the collision process.

Below, we present the way in which we have implemented the collision recipe with the grouping method of \citet{OrmelSpaans:2008}.

\subsection{Collision rates} 
The cycle starts with the calculation (or update) of the collision rates between the groups of the simulation.  The individual collision rate between two particles $i$ and $j$ is $C_{ij} = K_{ij}/{\cal V}$ (units: $\mathrm{s}^{-1}$), where ${\cal V}$ is the simulation volume and $K$ the collision kernel. For grouped collisions $C_{ij}$ is larger because many particles are involved in the collision.  The collision kernel $K$ is defined as $K_{ij}={\sigma}_{ij}^\mathrm{C} \Delta v_{ij}$ with $\sigma_{ij}^\mathrm{C} = \pi (a_\mathrm{out,1}+a_\mathrm{out,2})^2$ the collisional cross section (uncorrected for missing collisions) and $\Delta v_{ij}$ the average root-mean-square relative velocity (See \app{relvel}). Thus, to calculate the collision rates we need the relative velocities and the relation between the geometrical and the outer radius (\app{aout-asig}).

\subsection{Determination of collision partners} 
Random numbers determine which two groups collide and the number of particles that are involved from the $i$ and $j$ groups, $\eta_i$ and $\eta_j$. Then, each $i$-particle collides with $\eta_j/\eta_i$ $j$-particles. The grouping method implicitly assumes that collision rates do not change significantly during the collision process. To enforce the plausibility of this assumption the grouping method limits the total mass of the $j$-particles colliding with the $i$-particle to be at most 1\% of the mass of an $i$-particle, \ie\ $\eta_jm_j/\eta_im_i\lesssim f_\varepsilon = 10^{-2}$. Therefore, grouped collisions occur only in the local recipe.  For erosion or sticking this procedure works as intended. However, in collisions that result in breakage the grouping assumption is potentially problematic, since the particle properties -- and hence the collision rates -- then clearly change significantly over a single collision.  Fortunately, in the local recipe breakage is relatively unimportant.  Catastrophic disruptions (shattering) is problematic for the same reasons, because when it occurs, there is no `large' aggregate left. However, for energetic reasons we expect that shattering occurs mainly when two equal size particles are involved, for which the global recipe would apply (and no grouping).  In the following we continue with a collision of $\eta_t=\eta_j/\eta_i$ $j$-particles colliding with a single $i$-particle.

\begin{figure}
  \centering
  \includegraphics[width=80mm,clip]{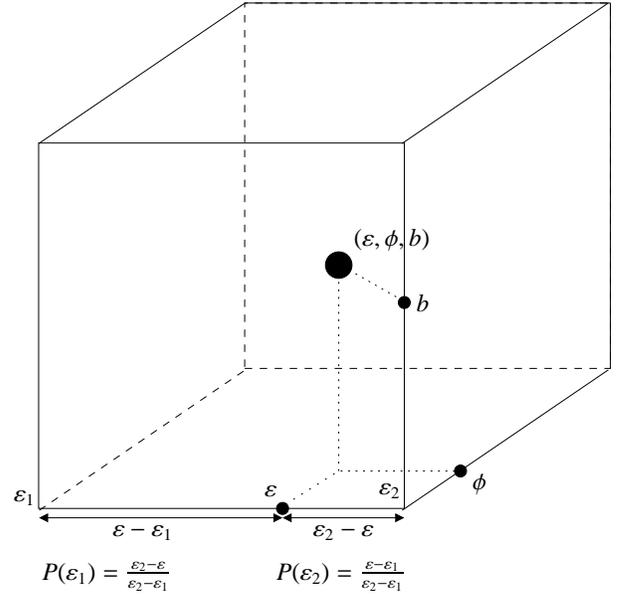}
  \caption{\label{fig:cube}Illustration of the picking of the grid points. The collision takes place at $(\varepsilon,\phi_\sigma,\tilde{b})$: a point that is generally surrounded by eight grid points (here corresponding to the nodes of the cube) at which results from the binary collision simulations are available. Each node is then assigned a probability inversely proportional to the distance to the grid point.  Thus, the probability that the energy parameter $\varepsilon=\varepsilon_1$ is picked (corresponding to four of the eight grid nodes) is $P_1=(\varepsilon-\varepsilon_1)/(\varepsilon_2-\varepsilon_1)$. The procedure is identical for the other grid points.}
\end{figure}
\subsection{Determining the collision quantities in grouped collisions} When the collision is in the `hit-and-stick' regime the properties of the new particles are easily found by adding the masses of the $j$-particles to the $i$-particle and calculating their filling factor using \eq{hs}. We therefore concentrate here on the local or global recipe. The collision is then characterized by the three dimensionless parameters: normalized collision energy $\varepsilon$, filling factor $\phi_\sigma$ and impact parameters $\tilde{b}$ (\se{par-space-normalization}).  
These three parameters constitute an arbitrary point in the 3D ($\varepsilon,\phi_\sigma,b$)-space, and will in general be confined by eight grid points ($k$) which correspond to the parameters at which results from the collision experiments are available, see \fg{cube}.  We next distribute the $\eta_t$ collisions over the grid point in which the weight of a grid point is inversely proportional to the `distance' to ($\varepsilon,\phi_\sigma,b$) as explained in \fg{cube}. Taking account of the collisions that result in a miss, we have
\begin{equation}
  \eta_t = \eta_\mathrm{miss} + \sum_{k=1}^8 \eta_k;\qquad \eta_\mathrm{miss}=\sum_{k=1}^8 \eta_{\mathrm{miss},k},
\end{equation}
where $\eta_{\mathrm{miss},k} \simeq \eta_t P_k f_{\mathrm{miss},k}$ denotes the number of collisions at the grid point resulting in a miss. Here, $P_k$ denotes the weight of the grid point ($\sum_k P_k =1$), $f_\mathrm{miss}$ the fraction of missed collisions at the grid point and the $\simeq$ sign indicates this number is rounded to integer values. Similarly, the number of `hits' at a grid point is given by $\eta_k \simeq \eta_t P_k (1-f_{\mathrm{miss},k})$. Not all of these grid points have to be occupied (\ie\ $\eta_k$ can be zero). In the special case without grouping $\eta_t=1$ and one grid point at most is occupied. 

We continue with the general case of multiple occupied grid points.  First, we consider the mass that is eroded, given by the $f_{\mathrm{pwl},k}$ quantities. The mass eroded at one grid point per collision is given by $M_{\mathrm{pwl},k} = f_{\mathrm{pwl},k} (m_i+m_j)$ (global recipe) or $M_{\mathrm{pwl},k} = f_{\mathrm{pwl},k} m_im_j/(m_i+m_j)$ (local recipe). Then, the total mass eroded by the group collision is
\begin{equation}
  M_\mathrm{pwl} = \sum_{k=1}^8 M_{\mathrm{pwl},k} \eta_k.
\end{equation}
If this quantity is more than $m_i$, clearly there is no large fragment component.\footnote{Recall that in grouped collisions ($\eta_\mathrm{t}>1$) this implies that the grouping method is not fully accurate as the change in mass is of the order of the mass itself; but the procedure is always fine if $\eta_\mathrm{t}=1$.} Otherwise, the mass of the large fragment component is $M_\mathrm{large} = m_i + (\eta_t-\eta_\mathrm{miss})m_j - M_\mathrm{pwl}$. Each $M_{\mathrm{pwl},k}$ quantity is distributed as a power-law with the exponent provided by the slope $q_k$ of the grid point (see below). Concerning the large-fragment component, there is a probability that it will break, given by the $N_{\mathrm{f},k}$ and $S_{\mathrm{f},k}$ quantities.  As argued before, breakage within the context of the grouping algorithm cannot be consistently modeled.  Notwithstanding these concerns, we choose to implement it in the grouping method. Because its probability is small, we assume it happens at most only once during the group collision. The probability that it occurs is then
\begin{equation}
  P_2 = 1 - \prod_{k=1}^8 (1-P_{2,k})^{\eta_k},
  \label{eq:P2}
\end{equation}
where $P_{2,k}$ is the probability that breakage occurs at a grid point and follows from the $S_\mathrm{f}$ and $N_\mathrm{f}$ quantities.  If breakage occurs, the masses $M_\mathrm{pwl}$ are removed first and we divide the remaining mass $M_\mathrm{large}$ in two.

The last quantity to determine is the change in the filling factor of the large aggregate, denoted by the $C_\phi$ symbol for one collision. Like \eq{P2} we multiply the changes in $C_\phi$ at the individual grid nodes,
\begin{equation}
  \phi_{\sigma,\mathrm{large}} = \langle \phi_\sigma \rangle_m \prod_{k=1}^8 C_{\phi,k}^{\eta_k},
  \label{eq:fCp}
\end{equation}
This completes the implementation of the collisional outcome within the framework of the grouping mechanism. That is, we have the masses and the filling factor of the large fragment component ($M_\mathrm{large}, \phi_{\sigma, \mathrm{large}}$), and have computed the distribution of the power-law component in terms of mass. Recall, finally, that all these results are \textit{per} $i$-particle, and that the multiplicity of the results is $\eta_i$.

\subsection{Picking of the power-law component masses} The final part of the MC cycle is to pick particles according to the power-law distribution, under the constraints of a total mass $\eta_kM_{\mathrm{pwl},k}$ and slope $q_k$ at each grid point $k$.  The general formula for picking the mass of the small fragments is
\begin{equation}
  m_\mathrm{small} = m_0\left[1 + \overline{r}\left( \left(\frac{m_\mathrm{rem}}{m_0}\right)^{1+q} -1\right)\right]^{1/(1+q)},
  \label{eq:mi}
\end{equation}
where $m_\mathrm{rem}$ is the remaining mass of the distribution (it starts at $m_\mathrm{rem} = \eta_kM_{\mathrm{pwl},k}$ and decreases every step by $m_\mathrm{small}$) and $\overline{r}$ a random number between 0 and 1.  From the definition of the power-law component $m_\mathrm{small}$ cannot be more than 25\% of the total mass.  In the MC program the number of \textit{distinct} fragments that can be produced is limited to a few per grid point.  This is to prevent an influx of a very large number of species (non-identical particles; in this case, particles of different mass), which would lead to severe computational problems, filling-up the statevector array (see below).  Therefore, if the same mass $m_\mathrm{small}$ is picked again it is considered to be the same species, and the multiplicity of this species is increased by one. After we have obtained a maximum of $\eta_\mathrm{dis}$ \textit{distinct} species, we redistribute the mass $M_\mathrm{pow}$ over the species.  In this way the fragment distribution is only sampled at a few discrete points.

\subsection{Merging/Duplication} The final part of the MC program consist of an inventory, and possible adjustment, of the amount of groups and species ($N_\mathrm{s}$) present in the program \citep{OrmelSpaans:2008}. To combine a sufficiently high resolution with an efficient computation in terms of speed is one of the virtues of the grouping method. One key parameter, determining the resolution of the simulation, is the $N_\mathrm{s}^\ast$ parameter (the target number of species in a simulation). In order to obtain a sufficient resolution we require that a total mass of $m_\mathrm{p}(t) N_\mathrm{s}^\ast$ is present in the simulation at all times, where $m_\mathrm{p}(t)$ is the mass peak of the distribution, $m_\mathrm{p}=\langle m^2 \rangle /\langle m \rangle$. Particles are \textit{duplicated} to fulfill this criterion, adding mass to the system. To prevent a pileup of species we followed the `equal mass method' \citep{OrmelSpaans:2008}. However, we found that due to the fragmentation many species were created at any rate -- too many, in fact ($N_\mathrm{s} > N_\mathrm{s}^\ast$) which would severely affect the efficiency of the program. Therefore, when $N_\mathrm{s}=2N_\mathrm{s}^\ast$ was reached we used the `merging algorithm' \citep{OrmelSpaans:2008} to combine neighboring species, a process that averages over their (structural) parameters but conserves mass. This significantly improved the efficiency (\ie\ speed) of the simulation, although the many fragments created by the collisions (all contributing to a higher $N_\mathrm{s}$) can be regarded as a redundancy, because it requires a lot of subsequent regrouping.  The alternative would be to produce only 1 new species per collision event \citep[see][]{ZsomDullemond:2008}.  Here, we prefer to stick with a more detailed representation of each collision event by creating a range of particles, but we acknowledge that this amount of detail is to some extent lost by the subsequent merging.
\section{List of symbols}
\label{app:appD}
  \begin{tabular}{p{8mm}p{80mm}}
   \hline
   \hline
   Symbol  & Description \\
   \hline
   ${\cal E^\ast}$           & Reduced modulus of elasticity  (\eq{Ebrroll})\\
   ${\cal R}_\mathrm{gd}$    & Gas-to-dust ratio by mass \\
   $\Delta v$                & Relative velocity (\app{app1})\\
   $\gamma$                  & Surface energy density (\eq{Ebrroll})\\
   $\eta$                    & Number of particles or groups (\app{mccycle})\\
   $\phi_\sigma$             & Geometrical filling factor (\eq{phi-geom-def})\\
   $\mu$                     & Molecular mass (\se{struc}) \\
   $\nu_\mathrm{m},\nu_\mathrm{t}$  & Molecular, turbulent viscosity (\se{working-model}\\
   $\xi_\mathrm{crit}$       & Critical displacement for irreversible rolling (\eq{Ebrroll}) \\
   $\rho_\mathrm{s}$         & Material density, $\rho_\mathrm{s}=2.65\ \mathrm{g\ cm^{-3}}$ (silicates) \\
   $\rho_\mathrm{g}$         & Gas density, $\rho_\mathrm{g} = \mu n m_\mathrm{H}$\\
   $\sigma$                  & Average projected surface area (\se{colmod}) \\
   $\sigma_{12}^\mathrm{C}$  & Collisional cross section (\se{colmod}) \\
   $\tau_\mathrm{f}$         & Friction time (\eq{tauf}) \\
   $C_\phi$                  & Change in geometrical filling factor, $C_\phi=\phi_\sigma/\phi_\sigma^\mathrm{ini}$ (\se{colrec}) \\
   $D_\mathrm{f}$            & Fractal dimension \\
   $E$                       & Collision energy, $E=\tfrac{1}{2} m_\mu (\Delta v)^2$ \\
   $E_\mathrm{roll}$         & Rolling energy (\eq{Ebrroll}) \\
   $E_\mathrm{br}$           & Breaking energy (\eq{Ebrroll}) \\
   $N$                       & Number of grains in aggregate (dimensionless measure of mass)\\
   $N_\mu$                   & Reduced number of monomers in collision $N_\mu = N_1N_2/(N_1+N_2)$ \\
   $N_\mathrm{f}$            & Number of big fragments\\
   $N_\mathrm{tot}$          & Total number of monomers in collision, $N_\mathrm{tot} = N_1 + N_2$\\
   $\mathrm{Re}$             & Reynolds number (\eq{Re})\\
   $S_\mathrm{f}$            & Spread in number of fragments of big component (\se{colrec})\\
   St                        & Particle Stokes number (\app{app1})\\
   $T$                       & Temperature (\se{working-model})\\
   $a_0$                     & Monomer radius \\
   $a_\mathrm{out}$          & Aggregate outer radius (\fg{fractal}) \\
   $a_\sigma$                & Aggregate geometrical radius (projected surface equivalent radius) (\fg{fractal})\\
   $a_\mu$                   & Reduced radius (\eq{Ebrroll})\\
   $b$                       & Impact parameter \\
   $b_\mathrm{max}$          & Sum of the outer radii, $b_\mathrm{max} = a_\mathrm{1,out}+a_\mathrm{2,out}$ \\
   $\tilde{b}$               & Normalized impact parameter, $\tilde{b}=b/b_\mathrm{max}$\\
   $c_\mathrm{g}$            & Sound speed (gas) \\
   $f_\mathrm{miss}$         & Fraction of collision misses (\se{colrec}) \\
   $f_\mathrm{pwl}$          & Fraction of mass in power-law component (\se{colrec}) \\
   $n$                       & Particle density (gas) \\
   $m$                       & Particle mass \\
   $m_\mu$                   & Reduced mass\\
   $m_\mathrm{H}$            & Hydrogen mass \\
   $q$                       & Power-law exponent (size distribution) (\se{colrec})\\
   $\overline{r}$            & Random deviate \\
   $t_\mathrm{ad}$           & Ambipolar diffusion time (\eq{tad}) \\
   $t_\mathrm{coll,0}$       & Initial collision time (\eq{tcoll}) \\
   $t_\mathrm{ff}$           & Free-fall time (\eq{tff}) \\
   $t_\mathrm{s}$            & Inner (Kolmogorov) eddy turn-over time (\eq{ts}) \\
   $v_\mathrm{L}$            & Large eddy turn-over velocity (\se{working-model})\\
   $v_\mathrm{s}$            & Inner (Kolmogorov) eddy turn-over velocity (\eq{vs})\\
   \hline
 \end{tabular}
\end{appendix}
\end{document}